\documentclass[conference,letterpaper,onecolumn]{IEEEtran}

\usepackage[utf8]{inputenc} 
\usepackage[T1]{fontenc}
\usepackage{url}
\usepackage{ifthen}
\usepackage{cite}
\usepackage[cmex10]{amsmath}
\usepackage{amssymb}

\usepackage{todonotes}

\newtheorem{theorem}{Theorem}
\newtheorem{proposition}[theorem]{Proposition}

\newtheorem{definition}[theorem]{Definition}

\newtheorem{lemma}[theorem]{Lemma}
\newtheorem{claim}[theorem]{Claim}

\newtheorem{remark}[theorem]{Remark}

\begin{document}

\title{The Rate-Distortion-Perception Trade-off:\\The Role of Private Randomness
}

\author{
  \IEEEauthorblockN{Yassine Hamdi\IEEEauthorrefmark{1}, Aaron B. Wagner\IEEEauthorrefmark{2}, Deniz G\"{u}nd\"{u}z\IEEEauthorrefmark{1}}
  \IEEEauthorblockA{\IEEEauthorrefmark{1}Department of Electrical and Electronic Engineering, Imperial College London, UK,}
  \IEEEauthorblockA{\{y.hamdi, d.gunduz\}@imperial.ac.uk}
  \IEEEauthorblockA{\IEEEauthorrefmark{2}School of Electrical and Computer Engineering, Cornell University, USA}
  \IEEEauthorblockA{wagner@cornell.edu}
}

\maketitle

\begin{abstract}
In image compression, with recent advances in generative modeling, the existence of a trade-off between the rate and the perceptual quality (realism) has been brought to light, where the realism is measured by the closeness of the output distribution to the source. It has been shown that randomized codes can be strictly better under a number of formulations.
In particular, the role of common randomness has been well studied.
We elucidate the role of private randomness in
the compression of a memoryless source $X^n=(X_1,...,X_n)$ under two kinds of realism constraints.
The
\textit{near-perfect realism} constraint requires the joint distribution of output symbols $(Y_1,...,Y_n)$ to be arbitrarily close the distribution of the source in total variation distance (TVD). The \textit{per-symbol near-perfect realism} constraint requires that the TVD between the distribution of output symbol $Y_t$ and the source distribution be arbitrarily small, uniformly in the index $t.$ We characterize the corresponding asymptotic rate-distortion trade-off and show that encoder private randomness is not useful if the compression rate is lower than the entropy of the source, however limited the resources in terms of common randomness and decoder private randomness may be.
\end{abstract}

\section{Introduction}
\label{sec:intro}

In conventional rate-distortion theory, the objective is to facilitate the reconstruction of a representation, denoted as $Y^n \triangleq (Y_1, ..., Y_n),$ of a source signal $X^n=(X_1, ..., X_n),$ while optimizing the proximity between the two, measured by a distortion measure $d(X^n,Y^n).$ The \textit{asymptotic regime} $n \to \infty$ has been extensively studied, starting with the work of Claude Shannon, who characterized the optimal asymptotic trade-off between rate and distortion, for additive distortion measures, i.e. $d(x^n,y^n) = (1/n)\sum_{t=1}^n d(x_t,y_t).$ The \textit{one-shot} scenario $n=1$ has also been studied \cite{2023BookFiniteBlocklengthSourceCoding}.
Despite the overall success of this theory (e.g.~\cite{Pearlman:Said,Sayood:Compression}), one notable limitation is the potential for the reconstructed output to manifest qualitatively distinct features from the original source realization. For a memoryless Gaussian source, when optimizing the mean-squared error (MSE) distortion measure, the reconstructed output typically possesses reduced power compared to the source.
Consequently, this phenomenon manifests as perceptual blurring in JPEG images at low bit-rates.
The concept of distortion measure only serves as a surrogate for the ultimate metric of genuine interest: how the reconstructed output is perceived by the end-user, typically a human observer. In certain scenarios, the latter may exhibit a preference for a reconstruction that registers higher distortion. A noteworthy example is MPEG Advanced Audio Coding (AAC): artificial noise is deliberately introduced into high-frequency bands~\cite[Sec.~17.4.2]{Sayood:Compression},
to align
the power spectrum of the reconstruction with that of the source.\\
\indent In conventional rate-distortion theory, it is known that deterministic encoders and decoders are sufficient to achieve the optimal asymptotic rate-distortion performance for a stationary source \textemdash as well as the optimal performance for one-shot fixed-length codes \cite{2003BookHanInfoSpectrum}.
The systematic study of the impact of perceptual quality constraints and randomization on rate and distortion took flight with the works of Li \textit{et al.} \cite{2010LiEtAlTheirFirstPaperOnDistributionPreservingQuantization,2011LiEtAlMainPaperOnDistributionPreservingQuantization,2012LiEtAlSpectralDensityPreservingQuantizationForAudio,2013LiEtAlMultipleDescriptionDistributionPreservingQuantization} and Saldi \textit{et al.} \cite{Jan2015SaldiEtAlDistributionPreservationMeasureTheoreticConsiderationsForContinuousAndDiscreteCommonRandomness,Sep2015RDPLimitedCommonRandomnessSaldi}. Therein, perceptual quality, or \textit{realism}, is formalized by requiring the distribution of the reconstruction $Y^n$ to be identical to that of the source, or asymptotically
arbitrarily close in total variation distance (TVD).
See also the work of Delp \textit{et al.}
\cite{1991MomentPreservingQuantization}.
Recently, in
\cite{2019AgustssonMentzerGANforExtremeCompression}, the authors used generative adversarial networks (GANs) to push the limits of image compression in very low bit-rates by synthesizing image content, such as facades of buildings, using a reference image database.
This line of work lead to the introduction \cite{2019BlauMichaeliRethinkingLossyCompressionTheRDPTradeoff} \textemdash see also \cite{2018PerceptionDistortionTradeOff,Aug2018MatsumotoRDPDeterministic,Nov2018MatsumotoRDPDeterministic}\textemdash \text{ }of a relaxed distribution-preservation constraint:
the problem is then to characterize the optimal rate for which both distortion constraint $d(X^n, Y^n) \leq \Delta,$ and realism constraint $\mathcal{D}(P_{X^n}, P_{Y^n}) \leq \lambda$ are met,
where
$\mathcal{D}$ 
is a similarity measure,
e.g., the TVD or some other divergence.
The three-way trade-off between $\Delta, \lambda,$ and the rate, was
coined
\textit{the rate-distortion-perception} (RDP) trade-off.
We call the above
the
\textit{strong realism constraint}, and \textit{imperfect strong realism} constraint when $\lambda>0.$
The following
weaker variant
has been recently studied \cite{Dec2022WeakAndStrongPerceptionConstraintsAndRandomness}:
\begin{IEEEeqnarray}{c}
\forall 1 \leq t \leq n, \quad \mathcal{D}(P_{Y_t}, p_X) \leq \lambda,\label{eq:def_imperfect_per_symbol_realism}
\end{IEEEeqnarray}
We call this \textit{
per-symbol realism}.
Other constraints depending directly on the realizations of the source and the reconstruction
have also been considered
\cite{Sep2015RDPLimitedCommonRandomnessSaldi,Dec2022WeakAndStrongPerceptionConstraintsAndRandomness,2023YangQiuAaronBWagnerUnifyingFidelityAndRealism}.\\
\indent The problem of characterizing the role of randomization under different formulations of the realism constraint, such as done in \cite{Sep2015RDPLimitedCommonRandomnessSaldi}, has very recently attracted renewed interest \cite{2022AaronWagnerRDPTradeoffTheRoleOfCommonRandomness,Dec2022WeakAndStrongPerceptionConstraintsAndRandomness} \textemdash see also \cite{2023XueyanGunduzISITConditionalRDP,OurISIT2023} when an additional source is available as side information, and \cite{UniversalRDPNeurips2021,2022JunChenKhistiBinarySourcesRDPAndFixedEncoderAndSuccessiveRefinement} for a successive refinement scenario. The different forms of randomness include private randomness at each of the encoder and decoder, and common randomness, available at both terminals. In the present work, we delve deeper into the role of private randomness.
We characterize the five-way trade-off between compression rate, common randomness rate, encoder private randomness rate, decoder private randomness rate and distortion, thereby extending previous results.
We consider a memoryless source $X^n \sim p_{X}^{\otimes n}$
and
the \textit{near-perfect strong realism}
and \textit{near-perfect per-symbol realism}
constraints:
\begin{IEEEeqnarray}{c}
    \|P_{Y^n} - p_{X}^{\otimes n}\|_{TV} \underset{n \to \infty}{\longrightarrow} 0\label{eq:_in_intro_def_near_perfect_realism}
\\
    \max_{1\leq t \leq n}\|P_{Y_t} - p_{X}\|_{TV} \underset{n \to \infty}{\longrightarrow} 0,\label{eq:_in_intro_def_per_symbol_near_perfect_realism}
\end{IEEEeqnarray}where $\|\cdot\|_{TV}$ is the
TVD.
We first introduce a novel soft covering result regarding the private randomness of stochastic compressors.
Then, we show that
whether encoder private randomness is available
does not impact the optimal asymptotic trade-off between rate and distortion. This holds whatever the resources in terms of common randomness and decoder private randomness are, as long as the compression rate is less than the entropy of the source.
This implies that in the absence of common randomness, it is not useful, in the limit of large blocklength $n,$ that the encoder include in its message a seed for a pseudo-random number generator. In other words, the only useful form of shared randomness is a common randomness available without communication.
\\
\indent The RDP trade-off has
strong ties to the channel simulation problem, a.k.a. reverse channel coding, a.k.a. channel synthesis, which can be stated \cite{2013PaulCuffDistributedChannelSynthesis} as that of finding the optimal rate such that
\begin{IEEEeqnarray}{c}
    \|P_{X^n,Y^n} - p_{X,Y}^{\otimes n}\|_{TV} \underset{n \to \infty}{\longrightarrow} 0,\nonumber
\end{IEEEeqnarray}for some target $p_{X,Y}.$
This problem has recently had successful applications in neural network based compression \cite{2020TheisAgustssonReIntroducingDitheredQuantization,TheisEtAlChannelSimulationWithDiffusionGaussian} and Federated Learning \cite{2023BurakGunduzChannelSimulationInFederatedLearning,2023DieuleveutHegazyChannelSimulationForFederatedLearning}.
Moreover, a channel simulation scheme was used to prove the first coding theorem \cite{TheisWagner2021VariableRateRDP} regarding the RDP trade-off with imperfect realism.
In the present work,
we
use
proof techniques from the channel simulation literature:
our proofs track those of \cite{2013PaulCuffDistributedChannelSynthesis}, and we use several of the soft covering lemma variants therein.
The paper is organized as follows. We give the problem formulation in Section \ref{sec:problem_formulation_and_equivalence}, and introduce a key lemma in Section \ref{sec:encoder_private_randomness}. We present all our other results in Section \ref{sec:main_result}, and provide a partial proof in Section \ref{sec:proofs_coding_theorems}. The rest of the proofs is provided in the appendices.

\section{Problem formulation}\label{sec:problem_formulation_and_equivalence}
\subsection{Notation}
\noindent Calligraphic letters such as $\mathcal{X}$ denote sets, except in $p^{\mathcal{U}}_{\mathcal{J}},$ which denotes the uniform distribution over alphabet $\mathcal{J}.$ 
Random variables are denoted using upper case letters such as $X,$ and their realizations using lower case letters such as $x.$
For a distribution $P,$ the expression $P_X$ denotes the marginal of variable $X,$ while $P(x)$ denotes the probability of the event $X{=}x.$ Similarly, $P_{X|Y{=}y}$ denotes a distribution over $\mathcal{X},$ and $P_{X|Y{=}y}(x)$ a real number.
We denote by $\mathbf{1}_{X{=}x_0}$ the distribution such that the events $X{=}x_0$ and $X{\neq}x_0$ have probabilities $1$ and $0.$
We denote by $[a]$ the set $\{1, ..., \lfloor a \rfloor\},$ and by $x^n$ the finite sequence $(x_1, ..., x_n).$  
The
TVD
between distributions $p$ and $q$ on a space $(\mathcal{X},\mathcal{B})$ is defined by
\begin{IEEEeqnarray}{c}
\|p-q\|_{TV} := \sup_{B \in \mathcal{B}} |p(B)-q(B)|.\nonumber
\end{IEEEeqnarray}
The closure of a set $\mathcal{A}$ is denoted by
$\overline{\mathcal{A}}.$ We use the definitions in \cite{2011GrayGeneralAlphabetsRelativeEntropy} for the entropy $H(X)$ and mutual information $I(X;Y)$ for random variables $X$ and $Y$ taking values in some Polish spaces.
This includes discrete spaces and real vector spaces. We always endow a Polish alphabet with the corresponding Borel $\sigma$-algebra. It contains all singletons. All conditional probability kernels we consider are regular.
We define $\overline{\mathbb{R}}_{\geq 0}{:=} \mathbb{R}_{\geq 0}{\cup}\{\infty\}$ and use the convention $\infty {\geq} \infty.$
The notation $\overset{\mathcal{P}}{\rightarrow}$ stands for convergence in probability.
For a finite alphabet 
$\mathcal{X},$
$\mathbb{P}^{\text{emp}}_{\mathcal{X}}(x^n)$ is the empirical distribution of
$x^n{\in}\mathcal{X}^n.$
Given a distribution $P_{X^n}$ on
$\mathcal{X}^n,$ we denote by $\hat{P}_{\mathcal{X}}[X^n]$
the \textit{average empirical distribution} of random string $X^n,$ i.e.,
the distribution on $\mathcal{X}$ defined by:
for any measurable $A {\subseteq} \mathcal{X},$
\begin{IEEEeqnarray}{c}
\scalebox{1.0}{$
\hat{P}_{\mathcal{X}}[X^n](A) = \tfrac{1}{n} \sum_{t=1}^n P_{X_t}(A)
$}
\nonumber
\end{IEEEeqnarray}

\subsection{Definitions}
\begin{figure}[t!]
    \centering\includegraphics[width=0.48\textwidth]{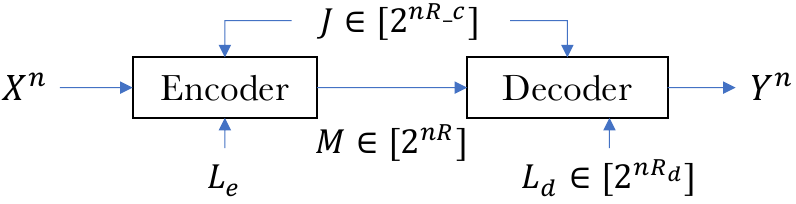}
    \caption{The system model.}
\label{fig:general_setup}
\end{figure}

\begin{definition}\label{def:distortion}
    Given a space $\mathcal{X},$ a distortion measure is a measurable function \scalebox{0.96}{$d:\mathcal{X}^2 {\to} [0,\infty)$} extending to sequences as $$d(x^n, y^n) = \tfrac{1}{n}\scalebox{0.95}{$\sum_{t=1}^n$} d(x_t,y_t).$$ 
\end{definition}
\noindent As shown in Figure \ref{fig:general_setup}, we consider
common randomness and
decoder private randomness available at rates $R_c$ and $R_d,$ respectively, which may be infinite. We consider both fixed-length and variable-length codes. Either the encoder private randomness is unconstrained, or it is completely unavailable.
\begin{definition}\label{def:code}
Given a
space $\mathcal{X}$
and a triplet
\scalebox{0.9}{$(R,R_c,R_d)\in\mathbb{R}_{\geq 0}\times\overline{\mathbb{R}}_{\geq 0}^2,$}
an \scalebox{0.9}{$(n, R, R_c,
R_d)$}
variable-length
code is a 
tuple \scalebox{0.9}{$(p_J,
p_{L_e},
p_{L_d},F^{(n)},G^{(n)})$}
consisting of 
distributions $p_J,
p_{L_e},
p_{L_d}$ on some Polish spaces,
a
deterministic
encoder
\begin{IEEEeqnarray}{c}
F^{(n)}: (X^n,J,L_e) \mapsto M
\end{IEEEeqnarray}with some Polish output space,
and a
deterministic
decoder
\begin{IEEEeqnarray}{c}
G^{(n)}: (M,J,L_d) \mapsto Y^n \in \mathcal{X}^n,
\quad \text{satisfying}
\\
R \geq
\tfrac{1}{n}H(M), \quad
R_c \geq
\tfrac{1}{n}H(J), \quad
R_d \geq
\tfrac{1}{n}H(L_d).
\nonumber
\end{IEEEeqnarray}
Given a distribution $p_{X}$ on $\mathcal{X},$
\textit{the distribution induced by the code}
is given by
\begin{IEEEeqnarray}{rCl}
P_{X^n, J,
L_e, 
M, L_d, Y^n} &:=& 
p_{X}^{\otimes n} \cdot p_J
\cdot p_{L_e}
\cdot p_{L_d} \cdot \mathbf{1}_{M{=}F^{(n)}(x^n,j,l_e)}
\nonumber\\*
&\cdot& 
\mathbf{1}_{Y^n{=}G^{(n)}(m,j,l_d)}.\nonumber
\end{IEEEeqnarray}
Note that if any of $R,R_c,R_d$ is null, then the corresponding random variable ($M$ or $J$ or $L_d$) is constant $P$-almost surely \textemdash see e.g. the discussion above \cite[Lemma~7.18]{2011GrayGeneralAlphabetsRelativeEntropy}.
Given a triplet \scalebox{0.9}{$(R,R_c,R_d)\in\mathbb{R}_{\geq 0}
^3
,
$}
an \scalebox{0.9}{$(n, R, R_c,
R_d)$}
\textit{fixed-length code} is
an variable-length code such that
\begin{IEEEeqnarray}{c}
M \in [2^{nR}] \text{ a.s.,}
\quad
p_J \equiv p^{\mathcal{U}}_{[2^{nR_c}]},
\quad
p_{L_d} 
\equiv p^{\mathcal{U}}_{[2^{nR_d}]}.
\nonumber
\end{IEEEeqnarray}
No constraint (alphabet, distribution, rate) is imposed on the encoder private randomness of a fixed-length (resp. variable-length) code.
Similarly, we define an \scalebox{0.9}{$(n,R,R_c,\infty)$} fixed-length code as
a variable-length code with $M \in [2^{nR}] \text{ a.s.}$ and $p_J \equiv p^{\mathcal{U}}_{[2^{nR_c}]}.$
An \scalebox{0.9}{$(n,R,R_c,R_d)$} fixed-length (resp. variable-length) \textit{code with non-privately randomized encoding} is an \scalebox{0.9}{$(n,R,R_c,R_d)$} fixed-length (resp. variable-length) code
for which $L_e$ is constant $P$-almost surely.
\end{definition}

\begin{definition}\label{def:achievability}
Consider a space $\mathcal{X},$ a distribution $p_{X}$ on $\mathcal{X},$
and a distortion measure $d.$
A tuple \scalebox{0.9}{$(R, R_c,
R_d, \Delta)\in\mathbb{R}_{\geq 0}\times\overline{\mathbb{R}}_{\geq 0}^3$} is said to be
achievable with \textit{near-perfect realism} with fixed-length (resp. variable-length) codes with privately randomized (resp. non-privately randomized) encoding if there exists a sequence
of \scalebox{0.9}{$(n, R, R_c,
R_d)$} fixed-length (resp. variable-length) codes with privately randomized (resp. non-privately randomized) encoding with induced distributions \scalebox{0.9}{$\{P^{(n)}\}_n$} such that 
\begin{IEEEeqnarray}{rCl}
&
\scalebox{1.0}{$
\limsup_{n \to \infty} \mathbb{E}_{P^{(n)}}
$}
\scalebox{1.0}{$
[d(X^n,Y^n)] \leq \Delta,
$}
&\label{eq:def_achievability_distortion_constraint}\\*
&\scalebox{0.99}{$
\|P^{(n)}_{Y^n} - p_{X}^{\otimes n}\|_{TV} \underset{n \to \infty}{\longrightarrow} 0.
$}
&\label{eq:def_achievability_near_perfect_realism_constraint}
\end{IEEEeqnarray}
For each of the above notions of achievability, we introduce the corresponding notion of achievability \textit{with per-symbol near-perfect realism}, defined by replacing \eqref{eq:def_achievability_near_perfect_realism_constraint} by
\begin{IEEEeqnarray}{c}
\scalebox{1.0}{$\max_{t\in[n]} \|P^{(n)}_{Y_t} - p_{X}\|_{TV} \underset{n \to \infty}{\longrightarrow} 0.$}\label{eq:def_achievability_per_symbol_realism_constraint}
\end{IEEEeqnarray}
The similar notions of achievability with \textit{perfect realism}
or
\textit{perfect per-symbol realism}
are defined by replacing \eqref{eq:def_achievability_near_perfect_realism_constraint} by:
\begin{IEEEeqnarray}{c}
\scalebox{1.0}{$\exists N \in \mathbb{N}, \forall n\geq N, \ P^{(n)}_{Y^n} \equiv p_{X}^{\otimes n}
\quad \text{or}
$}\label{eq:def_achievability_perfect_realism_constraint}
\\
\scalebox{1.0}{$\exists N \in \mathbb{N}, \forall n\geq N, \forall t\in[n], \ P^{(n)}_{Y_t} \equiv p_{X}.$}\label{eq:def_achievability_perfect_per_symbol_realism_constraint}
\end{IEEEeqnarray}
\end{definition}

\section{A soft covering lemma for randomized compressors}\label{sec:encoder_private_randomness}

\subsection{Statement}\label{sec:in_two_sided_achievability__removing_encoder_private_randomness}
\begin{proposition}\label{prop:removing_encoder_private_randomness}
Consider a finite input alphabet $\mathcal{X},$ a distribution $p_X$ on $\mathcal{X},$ non-negative reals $R,R_c$ and a sequence $\{F^{(n)}\}_{n \geq 1}$ of
encoders corresponding to a sequence of $(n,R,R_c,\infty)$ fixed-length codes.
The $n$-th induced distribution is denoted $P^{(n)}.$
\noindent If $R<H(X),$ then for any finite alphabet $\mathcal{V}$ and any sequence of deterministic mappings $\mathbf{v}^{(n)}:[2^{nR}]\times[2^{nR_c}] \to \mathcal{V}^n,$ there exists a sequence of deterministic maps
\begin{IEEEeqnarray}{c}
\scalebox{0.95}{$f^{(n)}: \mathcal{X}^n \times [2^{nR_c}]
\to [2^{nR}],
\quad \text{such that}
$}\nonumber
\\
\scalebox{0.9}{$\big\|\hat{\Tilde{P}}^{(n)}_{\mathcal{X}\times\mathcal{V}}[X^n, \mathbf{v}^{(n)}(M,J)] - \hat{P}^{(n)}_{\mathcal{X}\times\mathcal{V}}[X^n, \mathbf{v}^{(n)}(M,J)]\big\|_{TV} \underset{n \to \infty}{\longrightarrow} 0,$}
\nonumber
\\*
\scalebox{0.9}{$\big\|\Tilde{P}^{(n)}_{M,J} - P^{(n)}_{M,J}\big\|_{TV} \underset{n \to \infty}{\longrightarrow} 0
, \quad \text{where}
$}\label{eq:in_prop_encoder_private_randomness_TV_sur_M_J} \IEEEeqnarraynumspace
\\
\Tilde{P}^{(n)}_{X^n,J,M}:=p_{X}^{\otimes n} \cdot p^{\mathcal{U}}_{[2^{nR_c}]} \cdot \mathbf{1}_{M=f^{(n)}(X^n,J)}.\nonumber 
\end{IEEEeqnarray}
\end{proposition}

\noindent The result follows rather directly from applying
the soft covering lemma with a sequence of general sources and channels \cite[Corollary~VII.3]{2013PaulCuffDistributedChannelSynthesis}.
We provide a proof in Appendix \ref{app:use_of_soft_covering_general_source_and_channel}.
\vspace{5pt}
\begin{remark}\label{rem:memoryless_generator}
Consider the setting of Proposition \ref{prop:removing_encoder_private_randomness} and
let
$\{G^{(n)}\}_{n \geq 1}$
be the sequence of decoders in the initial codes.
Assume that
there exists a finite alphabet $\mathcal{V},$ a conditional probability $P_{Y|V}$ from $\mathcal{V}$ to $\mathcal{X},$ and a sequence $\{\mathbf{v}^{(n)}\}_{n \geq 1}$ of deterministic maps with $\mathbf{v}^{(n)}:[2^{nR}] \times [2^{nR_c}] \to \mathcal{V}^n,$ such that for any $(m,j)$ we have
\begin{IEEEeqnarray}{c}
\scalebox{0.95}{$G^{(n)}_{Y^n|M{=}m,J{=}j} \equiv \prod_{t{=}1}^n p_{Y|V{=}\mathbf{v}_t(m,j)}.$}
\end{IEEEeqnarray}
Then,
\textemdash e.g.
Lemma \ref{lemma:TV_same_channel}, Appendix \ref{app:TV_lemmas}\textemdash,
sequence \scalebox{0.98}{$\{f^{(n)}\}_{n {\geq} 1}$} of Proposition \ref{prop:removing_encoder_private_randomness} corresponding to $\{\mathbf{v}^{(n)}\}_{n \geq 1}$
satisfies
\begin{IEEEeqnarray}{c}
\scalebox{0.85}{$\big\|\hat{\Tilde{P}}^{(n)}_{\mathcal{X}^2\times\mathcal{V}}$}\scalebox{0.8}{$[X^n,Y^n, \mathbf{v}^{(n)}(M,J)]$} - \scalebox{0.85}{$\hat{P}^{(n)}_{\mathcal{X}^2\times\mathcal{V}}$}\scalebox{0.8}{$[X^n,Y^n, \mathbf{v}^{(n)}(M,J)]$}\scalebox{0.8}{$\big\|_{TV} \underset{n \to \infty}{\longrightarrow} 0$}
\nonumber
\\*
\scalebox{0.9}{$\big\|\Tilde{P}^{(n)}_{M,J,Y^n} - P^{(n)}_{M,J,Y^n}\big\|_{TV} \underset{n \to \infty}{\longrightarrow} 0
, \quad \text{where}
$}\label{eq:after_prop_encoder_private_randomness_TV_sur_Y} \IEEEeqnarraynumspace
\\
\scalebox{0.9}{$\Tilde{P}^{(n)}_{Y^n|M,J,X^n} \equiv P^{(n)}_{Y^n|M,J,X^n} := G^{(n)}_{Y^n|M,J}.$}\nonumber
\end{IEEEeqnarray}
\end{remark}
\vspace{5pt}


\section{The rate-distortion-perception trade-off with near-perfect realism}\label{sec:main_result}

\subsection{The role of encoder private randomness for finite source alphabets}
\indent We have the following characterization, which is an extension of \cite[Theorems~1~\&~5]{Sep2015RDPLimitedCommonRandomnessSaldi}.
The achievability is proved in Section \ref{sec:proofs_coding_theorems}, and the converse in Appendix \ref{app:converse_discrete_alphabets}.
\vspace{5pt}
\begin{theorem}\label{thm:region_discrete_alphabets}
Consider a finite source alphabet $\mathcal{X},$ a distribution $p_{X}$ on $\mathcal{X}$ 
and a distortion measure $d.$ 
Define the region $\mathcal{S}_D$
of $\scalebox{0.85}{$[0,H(X)) \times \overline{\mathbb{R}}_{\geq 0}^3$}$
as 
\begin{align}\label{eq:def_S_Discrete}
      & 
    \left\{ \begin{array}{rcl}
        \scalebox{0.9}{$(R, R_c, R_d, \Delta)$} &:& \scalebox{0.9}{$\exists \ p_{X,V,Y} \in \mathcal{D}_D,$}\\
        \scalebox{0.9}{$R$} &\geq& \scalebox{0.9}{$I_p(X;V)$} \\
        \scalebox{0.9}{$R+R_c$} &\geq& \scalebox{0.9}{$I_p(Y;V)$} \\
        \scalebox{0.9}{$R_d$} &\geq& \scalebox{0.9}{$H_p(Y|V)$} \\
        \scalebox{0.9}{$\Delta$} &\geq& \scalebox{0.9}{$\mathbb{E}_p[d(X, Y)]$}
    \end{array}\right\},
\end{align} with $\mathcal{D}_D$ defined as
\begin{align}\label{eq:def_D_Discrete}
       & 
    \left\{ \begin{array}{rcl}
        &p_{X,V,Y} : \scalebox{0.9}{$X \sim$ } p_{X}, \ p_{Y} \equiv p_{X}&\\
        &X - V - Y&\\
        &|\mathcal{V}| \leq |\mathcal{X}|^2+1&
    \end{array}\right\}.
\end{align}Denote by \scalebox{0.9}{$\mathcal{A}^{(f)}_D$} the set of \scalebox{0.9}{$(R,R_c,R_d,\Delta) \in [0,H(X)) \times \mathbb{R}_{\geq 0}^3$} achievable with near-perfect realism with fixed-length codes and by \scalebox{0.9}{$\mathcal{A}^{(u)}_D$} the set of tuples \scalebox{0.9}{$(R,R_c,R_d,\Delta) \in [0,H(X)) \times \overline{\mathbb{R}}_{\geq 0}^3$} achievable with near-perfect realism with variable-length codes.
Denote by \scalebox{0.9}{$\mathcal{A}^{(f,*)}_D$} the set of tuples \scalebox{0.9}{$(R,R_c,R_d,\Delta) \in [0,H(X)) \times \mathbb{R}_{\geq 0}^3$} achievable with near-perfect realism with non-privately randomized encoding and by \scalebox{0.9}{$\mathcal{A}^{(u,*)}_D$} the set of tuples \scalebox{0.9}{$(R,R_c,R_d,\Delta) \in [0,H(X)) \times \overline{\mathbb{R}}_{\geq 0}^3$} achievable with near-perfect realism with variable-length codes and non-privately randomized encoding. Then,
\begin{enumerate}
\item the aforementioned sets have identical closures in \scalebox{0.9}{$\overline{\mathbb{R}}_{\geq 0}^4:$}
\begin{IEEEeqnarray}{c}
\overline{\mathcal{A}^{(f)}_D}
=\overline{\mathcal{A}^{(u)}_D}
=\overline{\mathcal{A}^{(f,*)}_D}
=\overline{\mathcal{A}^{(u,*)}_D}
=\overline{S_D}.
\end{IEEEeqnarray}
\item the same holds if each notion of achievability is replaced by the corresponding achievability with no common randomness $(R_c=0)$ and $\mathcal{S}_D$ is replaced by its intersection with the hyperplane $R_c=0.$
\item the same holds if each notion of achievability is replaced by the corresponding achievability with no decoder private randomness $(R_d=0)$ and $\mathcal{S}_D$ is replaced by its intersection with the hyperplane $R_d=0.$
\end{enumerate}
\end{theorem}
\vspace{10pt}
Consequently, for lossy compression \scalebox{0.9}{$(R{<}H(X))$} with a near-perfect realism constraint, encoder private randomness is not useful whatever the available resources in terms of common randomness and decoder private randomness are. In particular, this holds even if the latter two sources of randomness are not rate-limited or not discrete.
It is unclear whether
encoder private randomness may
be useful
when the compression rate $R$ is greater than or equal to the entropy of the source.

\subsection{An extension to sources with infinite entropy}
Following \cite{2022AaronWagnerRDPTradeoffTheRoleOfCommonRandomness}, we use the following assumption in order to handle general alphabets.
\vspace{5pt}
\begin{definition}\label{def:uniform_integrability}\cite{2022AaronWagnerRDPTradeoffTheRoleOfCommonRandomness}
Given a space $\mathcal{X},$ a probability distribution $p$ on $\mathcal{X}$ and a distortion measure $d,$ we say that $(d,p)$ is uniformly integrable if and only if
\begin{equation*}
\forall \varepsilon>0,
\exists \tau>0, \quad
\sup_{\mathbb{P}_{X,Y,\xi}} \mathbb{E}[d(X,Y) \cdot \xi] \leq \varepsilon,
\end{equation*} where the supremum is taken over all distributions $\mathbb{P}_{X,Y,\xi}$ on $\mathcal{X}^2 \times \{0,1\}$ satisfying $\mathbb{P}_X \equiv \mathbb{P}_Y \equiv p$ and $\mathbb{P}(\{\xi=1\}) \leq \tau.$
\end{definition}
The property of Definition \ref{def:uniform_integrability} is satisfied if $\mathcal{X}$ is finite and $d$ does not take infinite values. It is also satisfied if $\mathcal{X} \subseteq \mathbb{R},$ $d$ is the MSE distortion measure, and $p_X$ has a finite second moment,
as proved in \cite[Appendix~E]{Dec2022WeakAndStrongPerceptionConstraintsAndRandomness}.
Our approach for turning a privately randomized encoder into a non-privately randomized one requires to work with finite alphabets. To that end, we introduce the following notion,
which involves a standard formalism for the notion of arbitrarily fine quantization. A \textit{quantizer} on a measurable space $\mathcal{X}$ is any measurable finite-valued map from $\mathcal{X}$ onto itself.
\vspace{5pt}
\begin{definition}\label{def:continuious_distortion_function}
Consider a
source alphabet $\mathcal{X},$
a $\sigma$-algebra $\mathcal{B}$
of subsets of $\mathcal{X},$
a probability distribution $p$ on $(\mathcal{X},\mathcal{B})$ and a distortion measure $d.$
We say that $(d,p)$ is \textit{quantizable} if
the following holds: 
there exists a sequence $\{\kappa^{(\ell)}\}_{\ell \in \mathbb{N}}$ of
quantizers of $\mathcal{X}$ such that
the corresponding partitions asymptotically generate $\mathcal{B},$ and 
for any $\tau,\varepsilon>0,$
\begin{IEEEeqnarray}{rCl}
& \exists B_\tau \in \mathcal{B}, \quad p(\mathcal{X} \setminus B_\tau) \leq \tau, \label{eq:in_def_quantizability_proba}&
\end{IEEEeqnarray}
and there exists $L_{\varepsilon, \tau}$ such that for any $\ell \geq L_{\varepsilon, \tau},$
\begin{IEEEeqnarray}{c}
\forall (x,y)\in B_\tau^2, \quad \big| d(x,y) - d\big(\kappa^{(\ell)}(x), \kappa^{(\ell)}(y)\big)\big| \leq \varepsilon,\label{eq:in_def_quantizability_distortion}
\\*
\forall (x,y)\in \mathcal{X}^2 \setminus B_\tau^2, \quad d\big(\kappa^{(\ell)}(x), \kappa^{(\ell)}(y)\big) \leq d(x,y) + \varepsilon.
\label{eq:in_def_quantizability_dealing_with_the_outside}
\IEEEeqnarraynumspace
\end{IEEEeqnarray}
\end{definition}
\vspace{5pt}
This property is satisfied if $\mathcal{X}$ is
finite,
and we have
\vspace{5pt}
\begin{claim}\label{claim:Euclidean_quantizable}
If $\mathcal{X}$
is a finite-dimensional real vector space and
$d$
denotes
the Euclidean distance, then for any distribution $p$ on $\mathcal{X}$ and any $s>0,$ tuple
$(d^s,p)$ is quantizable.
\end{claim}
We provide a proof in Appendix \ref{app:subsec:quantizability_Euclidean_space}.
We have the following characterization, which is an extension of \cite[Theorem~1]{Sep2015RDPLimitedCommonRandomnessSaldi} and \cite[Theorem~2]{2022AaronWagnerRDPTradeoffTheRoleOfCommonRandomness}. The proof is provided in Appendix \ref{app:proof_coding_thm_general}.
\vspace{5pt}
\begin{theorem}\label{thm:region_general_alphabets}
Consider a Polish source alphabet $\mathcal{X},$ a distribution $p_{X}$ on $\mathcal{X}$ having infinite entropy, 
and a distortion measure $d$ such that $(d,p_X)$ is uniformly integrable and quantizable. Define the region $\mathcal{S}_G$ as 
\begin{align}\label{eq:def_S_General}
      & 
    \left\{ \begin{array}{rcl}
        \scalebox{0.9}{$(R, R_c, \Delta) \in \mathbb{R}_{\geq 0}^3$} &:& \scalebox{0.9}{$\exists \ p_{X,V,Y} \in \mathcal{D}_G \text{ s.t. } $}\\
        \scalebox{0.9}{$R$} &\geq& \scalebox{0.9}{$I_p(X;V)$} \\
        \scalebox{0.9}{$R+R_c$} &\geq& \scalebox{0.9}{$I_p(Y;V)$} \\
        \scalebox{0.9}{$\Delta$} &\geq& \scalebox{0.9}{$\mathbb{E}_p[d(X, Y)]$}
    \end{array}\right\},
\end{align} with $\mathcal{D}_G$ defined as
\begin{align}\label{eq:def_D_General}
       & 
    \left\{ \begin{array}{rcl}
        &p_{X,V,Y} : \scalebox{0.9}{$X \sim$ } p_{X}, \ p_{Y} \equiv p_{X}&\\
        &X - V - Y&
    \end{array}\right\},
\end{align}where the alphabet of $V$ is constrained to be finite. Denote by $\mathcal{A}_G$ the set of triplets $(R,R_c, \Delta)\in \mathbb{R}_{\geq 0}^3$ such that $(R,R_c,\infty,\Delta)$ is achievable with near-perfect realism with fixed-length codes, and by $\mathcal{A}^{(f,*)}_G$ the set of triplets $(R,R_c, \Delta)\in \mathbb{R}_{\geq 0}^3$ such that $(R,R_c,\infty,\Delta)$ is achievable with near-perfect realism with fixed-length codes and non-privately randomized encoding.
Then, 
\begin{enumerate}
\item the aforementioned sets have identical closures in \scalebox{0.9}{$\mathbb{R}_{\geq 0}^3:$}
\begin{IEEEeqnarray}{c}
    \overline{\mathcal{A}^{(f)}_G} = \overline{\mathcal{A}^{(f,*)}_G} = \overline{S_G}.
\end{IEEEeqnarray}
\item the same holds if each notion of achievability is replaced by the corresponding achievability with no common randomness $(R_c=0)$ and $\mathcal{S}_G$ is replaced by its intersection with the hyperplane $R_c=0.$
\end{enumerate}
\end{theorem}
The assumption of unlimited decoder private randomness is not very restrictive as far as the study of encoder private randomness is concerned.
Indeed, for most general alphabets of interest, the total variation distance between a distribution having finite entropy and a distribution having infinite entropy is equal to $1.$ Hence, if the source distribution has infinite entropy, then achievability with near-perfect realism requires either $R_d$ or $R_c$ to be infinite (assuming $R$ is finite). Moreover, the case of unconstrained common randomness is trivial: there is no use for local randomness, since unlimited randomness can be extracted from the common randomness.
We conjecture that
the assumptions of
finite-valued auxiliary variable $V$ and fixed-length codes
are not restrictive.

\subsection{Per-symbol realism}

The following theorem is an extension of \cite[Theorem~4]{Dec2022WeakAndStrongPerceptionConstraintsAndRandomness}, which states that under a perfect per-symbol realism constraint, whether common randomness is available does not impact the optimal asymptotic trade-off between rate and distortion. We find that the same can be said of encoder private randomness, if
only 
near-perfect per-symbol realism is required.
\vspace{5pt}
\begin{theorem}\label{thm:region_per_symbol_realism_all_alphabets}
Theorems \ref{thm:region_discrete_alphabets} and \ref{thm:region_general_alphabets} also hold
if achievability with near-perfect realism is replaced by achievability with per-symbol near-perfect realism, and each lower bound on $R+R_c$ is removed in $\mathcal{S}_D$ and $\mathcal{S}_G.$
\end{theorem}
\noindent A proof is provided in Appendix \ref{app:proof_per_symbol_realism}.
Our result
also
complements
\cite[Theorem~2]{Dec2022WeakAndStrongPerceptionConstraintsAndRandomness},
which states that
under the per-symbol realism constraint \eqref{eq:def_imperfect_per_symbol_realism},
if $\lambda>0,$ then 
fully determinitic codes are sufficient to achieve the optimal asymptotic trade-off.\\

\section{Achievability proof for Theorem \ref{thm:region_discrete_alphabets}}\label{sec:proofs_coding_theorems}

\subsection{Modifying a standard code construction}

A rather straightforward adaptation of the proof of \cite[Theorem~2]{2022AaronWagnerRDPTradeoffTheRoleOfCommonRandomness} yields the following result.
\vspace{5pt}
\begin{proposition}\label{prop:can_construct_a_discrete_code}
Consider finite alphabets \scalebox{0.9}{$\mathcal{X},\mathcal{V},$} a distortion measure \scalebox{0.9}{$d$} on \scalebox{0.9}{$\mathcal{X}^2,$} a triplet \scalebox{0.9}{$(R,R_c,\Delta)\in\mathbb{R}_{\geq 0}^3,$} and a distribution \scalebox{0.9}{$p_{X,Y,V}$} on \scalebox{0.9}{$\mathcal{X}^2\times\mathcal{V}.$} Assume that \scalebox{0.9}{$p \in \mathcal{D}_D$} and
\begin{IEEEeqnarray}{c}
\scalebox{1.0}{$R \geq I_p(X;V), \
R+R_c \geq I_p(Y;V), \
\Delta \geq \mathbb{E}_p[d(X,Y)].$}\nonumber
\end{IEEEeqnarray}Then, there exists a sequence \scalebox{0.9}{$\{\varepsilon_n\}_{n \geq 1}$} of positive reals
and a sequence of \scalebox{0.9}{$(n,R+\varepsilon_n,R_c,\infty)$} fixed-length codes inducing distributions \scalebox{0.9}{$\{P^{(n)}_{X^n,J,M,Y^n}\}_{n
\geq 1
}$} such that
\scalebox{0.9}{$\varepsilon_n \underset{n \to \infty}{\rightarrow} 0$} and
\begin{IEEEeqnarray}{c}
\scalebox{0.9}{$\limsup_{n \to \infty} \mathbb{E}_{P^{(n)}}[d(X^n,Y^n)] \leq \Delta$} \label{eq:distortion_on_P_we_get_from_lemma_discrete_code}\\
\scalebox{0.9}{$\|P^{(n)}_{Y^n}-p_X^{\otimes n}\|_{TV} \underset{n \to \infty}{\longrightarrow} 0$}\label{eq:percetion_we_get_on_P_from_lemma_discrete_code}\\
\scalebox{0.9}{$\forall (m,j), \ P^{(n)}_{Y^n|M{=}m,J{=}j} = \prod_{t=1}^n p_{Y|V{=}v^{(n)}_t(m,j)},$}\label{eq:decoder_is_a_memoryless_generator_from_lemma_discrete_code}\\
\scalebox{0.9}{$P^{(n)}\big(\big\|\mathbb{P}^{\text{emp}}_{\mathcal{V}}(v^{(n)}(M,J)) - p_V\big\|_{TV} \geq \varepsilon_n\big) \underset{n \to \infty}{\longrightarrow} 0,$}\label{eq:codeword_letter_frequency_converges_in_proba_from_lemma_discrete_code}
\end{IEEEeqnarray}for some sequence of deterministic functions \scalebox{0.9}{$v^{(n)}: [2^{n(R+\varepsilon)}]{\times}[2^{nR_c}] \to \mathcal{V}^{n}.$}
\end{proposition}
\noindent See Appendix \ref{app:subsec:proof_can_construct_a_discrete_code} for a proof.
As a direct consequence of Proposition \ref{prop:removing_encoder_private_randomness} and Remark \ref{rem:memoryless_generator}, we can state the following.
\vspace{5pt}
\begin{proposition}\label{prop:can_construct_non_privately_randomized_discrete_code}
For $R<H(X),$ Proposition \ref{prop:can_construct_a_discrete_code} holds with $P^{(n)}$ induced by a \scalebox{0.9}{$(n,R+\varepsilon_n,R_c,\infty)$} fixed-length code with non-privately randomized encoding, for large enough $n.$
\end{proposition}

\subsection{Achievability of Theorem \ref{thm:region_discrete_alphabets}}
By definition, for each of the three statements of Theorem \ref{thm:region_discrete_alphabets}, we know that $\mathcal{A}^{(f)}_D,\mathcal{A}^{(u)}_D,\mathcal{A}^{(u,*)}_D$ contain $\mathcal{A}^{(f,*)}_D.$ In this section, we consider the (more general) setting of the first statement and prove that $\mathcal{S}_D {\subseteq} \overline{\mathcal{A}^{(f,*)}_D}.$ Throughout the section, we explain how our proof implies that the same holds in the respective settings of the two other statements.
Let $(R,R_c,R_d,\Delta)$ be a tuple in $\mathcal{S}_D$ and $p_{X,V,Y}$ be a corresponding distribution in $\mathcal{D}_D.$ If $R_c = \infty,$ we replace it by $H(X),$ if $R_d =\infty,$ we replace it with $H(Y|X),$ and if $\Delta=\infty,$ we replace it with $\max(d),$ which exists since $\mathcal{X}$ is finite and $d$ is assumed to only take finite values (Definition \ref{def:distortion}). Then, \scalebox{0.9}{$(R,R_c,R_d,\Delta) \in [0,H(X))\times\mathbb{R}_{\geq 0}^3,$} and we can apply Proposition \ref{prop:can_construct_non_privately_randomized_discrete_code}.
 Hereafter, we use the notation of Proposition \ref{prop:can_construct_non_privately_randomized_discrete_code}.
If $R_d=0,$ then $H_p(Y|V)=0,$ thus from \eqref{eq:decoder_is_a_memoryless_generator_from_lemma_discrete_code}, for any $n \in \mathbb{N},$ $P^{(n)}$ is the distribution induced by a $(n,R+\varepsilon_n,R_c,0)$ fixed-length code. Hence, under the setting of the third statement of Theorem \ref{thm:region_discrete_alphabets}, we have $\mathcal{S}_D {\subseteq} \overline{\mathcal{A}^{(f,*)}_D},$ which concludes the proof regarding that setting. Moving to the case $R_d>0,$
we can use the same argument as in \cite[Appendix~2]{2013PaulCuffDistributedChannelSynthesis} and reach the following claim \textemdash see Appendix \ref{app:subsec:local_channel_synthesis} for a proof.
\vspace{5pt}
\begin{claim}\label{claim:local_channel_synthesis}
From the local channel synthesis lemma \cite[Corollary~VII.6]{2013PaulCuffDistributedChannelSynthesis}, \eqref{eq:decoder_is_a_memoryless_generator_from_lemma_discrete_code} and \eqref{eq:codeword_letter_frequency_converges_in_proba_from_lemma_discrete_code},
for any $\gamma>0$ and any $n \in \mathbb{N},$
there exists a mapping $\Tilde{P}_{Y^n|V^n}$ with a fixed-length private randomness of rate 
$H_p(Y|V)+\gamma$
such that
\begin{IEEEeqnarray}{c}
\scalebox{0.95}{$\big\|P^{(n)}_{X^n,M,J,v^n(M,J)} \cdot \Tilde{P}_{Y^n|V^n} - P^{(n)}_{X^n,M,J,v^n(M,J),Y^n}\big\|_{TV} 
\underset{n\to\infty}{\longrightarrow}0.$}\nonumber
\end{IEEEeqnarray}
\end{claim}
Hence, from Lemmas \ref{lemma:TV_joint_to_TV_marginal} and \ref{lemma:continuity_TV}, and since $d$ is bounded, replacing \scalebox{0.9}{$P^{(n)}_{Y^n|V^n}$} by \scalebox{0.9}{$\Tilde{P}_{Y^n|V^n}$} for every $n$ preserves
\eqref{eq:distortion_on_P_we_get_from_lemma_discrete_code}
and 
\eqref{eq:percetion_we_get_on_P_from_lemma_discrete_code},
and results in a \scalebox{0.9}{$(n,R+\varepsilon_n,R_c,R_d+\gamma)$} fixed-length code.
This holds for every $\gamma>0.$ Since the common randomness rate used is precisely $R_c,$ both when $R_d=0$ and when $R_d>0,$ this shows that we have \scalebox{0.9}{$\mathcal{S}_D \in \overline{A^{(f,*)}_D}$} in both the settings of the first and second statements of Theorem \ref{thm:region_discrete_alphabets}, which concludes the proof.


\section{Conclusion}
\label{sec:conclusion}

We have studied the role of private randomness in the rate-distortion-perception trade-off with near-perfect realism and near-perfect per-symbol realism constraints, in the classical infinite blocklength scenario. Our work complements previous results on the key role of randomization in these settings.
We have characterized the corresponding rate-distortion-perception trade-offs under different situations in terms of the amount of common and private randomness available.
Our results show that encoder private randomness is not useful when the compression rate is lower than the entropy of the source.
In particular, in that case, if no common randomness is available, it is not useful to send randomness from the encoder.
A similar phenomenon was conjectured \cite{2013PaulCuffDistributedChannelSynthesis} to hold for the channel synthesis problem, but this
has not yet been proved.
The role of encoder private randomness in the finite-blocklength compression setting merits further investigation.

\enlargethispage{-1.4cm} 

\section*{Acknowledgments}
%
The present work has received funding from the European Union’s Horizon 2020 Marie Sk\l{}odowska Curie Innovative Training Network Greenedge (GA. No. 953775).

\newpage
\bibliographystyle{IEEEtran}
\bibliography{biblio}

\begin{thebibliography}{10}
\providecommand{\url}[1]{#1}
\csname url@samestyle\endcsname
\providecommand{\newblock}{\relax}
\providecommand{\bibinfo}[2]{#2}
\providecommand{\BIBentrySTDinterwordspacing}{\spaceskip=0pt\relax}
\providecommand{\BIBentryALTinterwordstretchfactor}{4}
\providecommand{\BIBentryALTinterwordspacing}{\spaceskip=\fontdimen2\font plus
\BIBentryALTinterwordstretchfactor\fontdimen3\font minus \fontdimen4\font\relax}
\providecommand{\BIBforeignlanguage}[2]{{%
\expandafter\ifx\csname l@#1\endcsname\relax
\typeout{** WARNING: IEEEtran.bst: No hyphenation pattern has been}%
\typeout{** loaded for the language `#1'. Using the pattern for}%
\typeout{** the default language instead.}%
\else
\language=\csname l@#1\endcsname
\fi
#2}}
\providecommand{\BIBdecl}{\relax}
\BIBdecl

\bibitem{2023BookFiniteBlocklengthSourceCoding}
L.~Zhou and M.~Motani, \emph{{Finite Blocklength Lossy Source Coding for Discrete Memoryless Sources}}.\hskip 1em plus 0.5em minus 0.4em\relax {New Foundations and Trends}, 2023.

\bibitem{Pearlman:Said}
W.~A. Pearlman and A.~Said, \emph{{Digital Signal Compression: Principles and Practice}}.\hskip 1em plus 0.5em minus 0.4em\relax {Cambridge (England)}: {Cambridge University Press}, 2011.

\bibitem{Sayood:Compression}
K.~Sayood, \emph{{Introduction to Data Compression}}, 4th~ed.\hskip 1em plus 0.5em minus 0.4em\relax {Waltham, MA (United States of America)}: {Morgan Kaufmann}, 2012.

\bibitem{2003BookHanInfoSpectrum}
T.~S. Han, \emph{{Information-Spectrum Methods in Information Theory}}, {English}~ed.\hskip 1em plus 0.5em minus 0.4em\relax {Berlin, Heidelberg (Germany)}: {Springer}, 2003.

\bibitem{2010LiEtAlTheirFirstPaperOnDistributionPreservingQuantization}
M.~Li, J.~Klejsa, and W.~B. Kleijn, ``{Distribution Preserving Quantization With Dithering and Transformation},'' \emph{{IEEE Signal Processing Letters}}, vol.~17, no.~12, 2010.

\bibitem{2011LiEtAlMainPaperOnDistributionPreservingQuantization}
M.~Li, J.~Klejsa, and W.~Kleijn, ``{On Distribution Preserving Quantization},'' 2011, arXiv:1108.3728.

\bibitem{2012LiEtAlSpectralDensityPreservingQuantizationForAudio}
M.~Li, J.~Klejsa, A.~Ozerov, and W.~B. Kleijn, ``{Audio coding with power spectral density preserving quantization},'' in \emph{{IEEE International Conference on Acoustics, Speech and Signal Processing}}, 2012.

\bibitem{2013LiEtAlMultipleDescriptionDistributionPreservingQuantization}
J.~Klejsa, G.~Zhang, M.~Li, and W.~B. Kleijn, ``{Multiple Description Distribution Preserving Quantization},'' \emph{{IEEE Transactions on Signal Processing}}, vol.~61, no.~24, 2013.

\bibitem{Jan2015SaldiEtAlDistributionPreservationMeasureTheoreticConsiderationsForContinuousAndDiscreteCommonRandomness}
N.~Saldi, T.~Linder, and S.~Y{\"u}ksel, ``{Randomized Quantization and Source Coding With Constrained Output Distribution},'' \emph{{IEEE Transactions on Information Theory}}, vol.~61, no.~1, 2015.

\bibitem{Sep2015RDPLimitedCommonRandomnessSaldi}
------, ``{Output Constrained Lossy Source Coding With Limited Common Randomness},'' \emph{{IEEE Transactions on Information Theory}}, vol.~61, no.~9, 2015.

\bibitem{1991MomentPreservingQuantization}
E.~J. Delp and O.~R. Mitchell, ``{Moment preserving quantization (signal processing)},'' \emph{{IEEE Transactions on Communications}}, vol.~39, no.~11, 1991.

\bibitem{2019AgustssonMentzerGANforExtremeCompression}
E.~Agustsson, M.~Tschannen, F.~Mentzer, R.~Timofte, and L.~V. Gool, ``{Generative Adversarial Networks for Extreme Learned Image Compression},'' in \emph{{IEEE/CVF International Conference on Computer Vision, 2019}}.

\bibitem{2019BlauMichaeliRethinkingLossyCompressionTheRDPTradeoff}
Y.~Blau and T.~Michaeli, ``{Rethinking Lossy Compression: The Rate-Distortion-Perception Tradeoff},'' in \emph{{36th International Conference on Machine Learning}}, 2019.

\bibitem{2018PerceptionDistortionTradeOff}
------, ``{The Perception-Distortion Tradeoff},'' in \emph{{IEEE/CVF Conference on Computer Vision and Pattern Recognition, 2018}}.

\bibitem{Aug2018MatsumotoRDPDeterministic}
R.~Matsumoto, ``{Introducing the perception-distortion tradeoff into the rate-distortion theory of general information sources},'' \emph{{IEICE Communications Express}}, vol.~7, no.~11, 2018.

\bibitem{Nov2018MatsumotoRDPDeterministic}
------, ``{Rate-distortion-perception tradeoff of variable-length source coding for general information sources},'' \emph{{IEICE Communications Express}}, vol.~8, no.~2, 2019.

\bibitem{Dec2022WeakAndStrongPerceptionConstraintsAndRandomness}
J.~Chen, L.~Yu, J.~Wang, W.~Shi, Y.~Ge, and W.~Tong, ``{On the Rate-Distortion-Perception Function},'' \emph{{IEEE Journal on Selected Areas in Information Theory}}, vol.~3, no.~4, 2022.

\bibitem{2023YangQiuAaronBWagnerUnifyingFidelityAndRealism}
Y.~Qiu, A.~B. Wagner, J.~Ball\'{e}, and L.~Theis, ``{Wasserstein Distortion: Unifying Fidelity and Realism},'' 2023, arXiv:2310.03629.

\bibitem{2022AaronWagnerRDPTradeoffTheRoleOfCommonRandomness}
A.~B. Wagner, ``{The Rate-Distortion-Perception Tradeoff: The Role of Common Randomness},'' 2022, arXiv:2202.04147.

\bibitem{2023XueyanGunduzISITConditionalRDP}
X.~Niu, D.~G{\"u}nd{\"u}z, B.~Bai, and W.~Han, ``{Conditional Rate-Distortion-Perception Trade-Off},'' in \emph{{IEEE International Symposium on Information Theory}}, 2023.

\bibitem{OurISIT2023}
Y.~Hamdi and D.~G\"{u}nd\"{u}z, ``{The Rate-Distortion-Perception Trade-off with Side Information},'' in \emph{{IEEE International Symposium on Information Theory}}, 2023.

\bibitem{UniversalRDPNeurips2021}
G.~Zhang, J.~Qian, J.~Chen, and A.~Khisti, ``{Universal Rate-Distortion-Perception Representations for Lossy Compression},'' in \emph{{35th Annual Conference on Neural Information Processing Systems}}, 2021.

\bibitem{2022JunChenKhistiBinarySourcesRDPAndFixedEncoderAndSuccessiveRefinement}
J.~Qian, G.~Zhang, J.~Chen, and A.~Khisti, ``{A Rate-Distortion-Perception Theory for Binary Sources},'' in \emph{{International Zurich Seminar on Information and Communication}}, 2022.

\bibitem{2013PaulCuffDistributedChannelSynthesis}
P.~Cuff, ``{Distributed Channel Synthesis},'' \emph{{IEEE Transactions on Information Theory}}, vol.~59, no.~11, 2013.

\bibitem{2020TheisAgustssonReIntroducingDitheredQuantization}
E.~Agustsson and L.~Theis, ``{Universally Quantized Neural Compression},'' in \emph{34th Annual Conference on Neural Information Processing Systems}, 2020.

\bibitem{TheisEtAlChannelSimulationWithDiffusionGaussian}
L.~Theis, T.~Salimans, M.~D. Hoffman, and F.~Mentzer, ``Lossy compression with gaussian diffusion,'' 2022, arXiv.2206.08889.

\bibitem{2023BurakGunduzChannelSimulationInFederatedLearning}
B.~{Hasircioglu} and D.~{Gunduz}, ``{Communication Efficient Private Federated Learning Using Dithering},'' 2023, arXiv:2309.07809.

\bibitem{2023DieuleveutHegazyChannelSimulationForFederatedLearning}
M.~{Hegazy}, R.~{Leluc}, C.~T. {Li}, and A.~{Dieuleveut}, ``{Compression with Exact Error Distribution for Federated Learning},'' 2023, arXiv:2310.20682.

\bibitem{TheisWagner2021VariableRateRDP}
L.~Theis and A.~B. Wagner, ``{A coding theorem for the rate-distortion-perception function},'' in \emph{{Neural Compression: From Information Theory to Applications -- workshop at the International Conference on Learning Representations 2021}}.

\bibitem{2011GrayGeneralAlphabetsRelativeEntropy}
R.~M. Gray, \emph{{Entropy and Information Theory}}, 2nd~ed.\hskip 1em plus 0.5em minus 0.4em\relax New York, NY (United States of America): {Springer}, 2011.

\end{thebibliography}

\newpage

\appendices

\section{Some lemmas on the total variation distance}\label{app:TV_lemmas}
\begin{lemma}\label{lemma:TV_joint_to_TV_marginal}\cite[Lemma~V.1]{2013PaulCuffDistributedChannelSynthesis}
    Let $\Pi$ and $\Gamma$ be two distributions on an alphabet $\mathcal{W} \times \mathcal{L}.$ Then \begin{equation*}
        \| \Pi_W - \Gamma_W \|_{TV} \leq \| \Pi_{W,L} - \Gamma_{W,L} \|_{TV}.
    \end{equation*}
\end{lemma}
\begin{lemma}\label{lemma:TV_same_channel}\cite[Lemma~V.2]{2013PaulCuffDistributedChannelSynthesis}
    Let $\Pi$ and $\Gamma$ be two distributions on an alphabet $\mathcal{W} \times \mathcal{L}.$ Then when using the same channel $\Pi_{L|W}$ we have \begin{equation*}
        \| \Pi_W \Pi_{L|W} - \Gamma_W \Pi_{L|W} \|_{TV} = \| \Pi_W - \Gamma_W \|_{TV}.
    \end{equation*}
\end{lemma}
\begin{lemma}\label{lemma:get_expectation_out_of_TV}
    Let $\Pi$ be two distributions on the product of two Polish spaces $\mathcal{W} $ and $\mathcal{L},$ and let $\Pi_{L|W}, \Gamma_{L|W}$ be two channels. Then, we have \begin{equation*}
        \| \Pi_W \Pi_{L|W} - \Pi_W \Gamma_{L|W} \|_{TV} = \mathbb{E}_{\Pi_W} \big[ \| \Pi_{L|W} - \Gamma_{L|W} \|_{TV} \big].
    \end{equation*}
\end{lemma}
\begin{lemma}\label{lemma:continuity_TV}
Let $\Pi$ and $\Gamma$ be two distributions on a set $\mathcal{W},$ and $f:\mathcal{W} \to \mathbb{R}$ be a bounded function. Then,
\begin{IEEEeqnarray}{c}
| \ \mathbb{E}_{\Pi}[f] - \mathbb{E}_{\Gamma}[f] \ |
\leq
2\max |f| \cdot \|\Pi-\Gamma\|_{TV}.
\nonumber
\end{IEEEeqnarray}
\end{lemma}

\section{Proof of Proposition \ref{prop:removing_encoder_private_randomness}}\label{app:use_of_soft_covering_general_source_and_channel}
We use the soft covering lemma with a sequence of general sources and channels \cite[Corollary~VII.3]{2013PaulCuffDistributedChannelSynthesis}, which we state for completeness. For any distribution $\Phi_{V,W},$ we use the notation
\begin{IEEEeqnarray}{c}
i_\Phi(v;w) = \log(\Phi(v,w)/\Phi(v)\Phi(w))
\nonumber\\
i_\Phi(w) = -\log(\Phi(w)).\nonumber
\end{IEEEeqnarray}
\begin{lemma}\cite[Corollary~VII.3]{2013PaulCuffDistributedChannelSynthesis}\label{lemma:soft_covering_general_source_and_channel}
Let $\{\mathcal{U}^{(n)}, \mathcal{V}^{(n)}, \mathcal{W}^{(n)}\}_{n\geq 1}$ be a sequence of finite alphabets
and
$\{\Phi_{U^{(n)},V^{(n)},W^{(n)}}\}_{n {\geq} 1}$ be a sequence of distributions, the $n$-th being on $\mathcal{U}^{(n)}{\times}\mathcal{V}^{(n)}{\times}\mathcal{W}^{(n)}$.
For every $n {\geq} 1,$ and every $w^{(n)} {\in} \mathcal{W}^{(n)},$ let $u^{(n)}(w^{(n)})$ be a random variable with distribution $\Phi_{U^{(n)}|W^{(n)}=w^{(n)}}.$
Denote the family $\{u^{(n)}(w^{(n)})\}_{w^{(n)} {\in} \mathcal{W}^{(n)}}$ by $\mathcal{B}^{(n)}.$
For every $n {\geq} 1,$ define
\begin{IEEEeqnarray}{c}
\scalebox{0.98}{$\Tilde{\Phi}_{W^{(n)},U^{(n)},V^{(n)}} := \Phi_{W^{(n)}} \cdot \mathbf{1}_{U^{(n)}=u^{(n)}(w^{(n)})} \cdot \Phi_{V^{(n)}|W^{(n)},U^{(n)}}.$}\nonumber
\end{IEEEeqnarray}
Assume that
\begin{IEEEeqnarray}{c}
\scalebox{0.9}{$i_\Phi(W^{(n)},U^{(n)};V^{(n)})- i_\Phi(W^{(n)}) \overset{\mathcal{P}}{\longrightarrow} -\infty.$}\label{eq:assumption_in_soft_covering_general_source_and_channel}\IEEEeqnarraynumspace
\end{IEEEeqnarray}Then, we have
\begin{IEEEeqnarray}{c}
\mathbb{E}_{\mathcal{B}^{(n)}}[\|\Tilde{\Phi}_{V^{(n)}}-\Phi_{V^{(n)}}\|_{TV}] \underset{n\to\infty}{\longrightarrow} 0.\label{eq:TV_goes_to_zero_in_soft_covering_lemma_general_source_and_channel}
\end{IEEEeqnarray}
\end{lemma}
For every $n{\geq}1$
let $M{=}\psi^{(n)}(X^n,J,U)$ be a functional representation of $F^{(n)}.$
Take $\Phi_{W^{(n)},U^{(n)},V^{(n)}}$ corresponding to $P^{(n)}$ with $W^{(n)}{=}(X^n,J),$ $U^{(n)}{=}U,$ and $V^{(n)}{=}(M,J,\mathbb{P}^{\text{emp}}_{X^n,\mathbf{v}^{(n)}(M,J)}).$\\

\noindent \underline{Proving \eqref{eq:assumption_in_soft_covering_general_source_and_channel}}
\hfill\\
Since $V^{(n)}$ is a deterministic function of $(W^{(n)},U^{(n)}),$
it can be easily checked that $1/\Phi_{V^{(n)}}$ is a density for $\Phi_{W^{(n)},U^{(n)},V^{(n)}}$ with respect to $\Phi_{W^{(n)},U^{(n)}}\cdot \Phi_{V^{(n)}}.$
Consider \scalebox{0.9}{$\varepsilon_1,\varepsilon_2 \in (0,H(X){-}R)$} such that \scalebox{0.9}{$\varepsilon_1+\varepsilon_2<H(X){-}R.$}
\begin{IEEEeqnarray}{rCl}
\IEEEeqnarraymulticol{3}{l}{
P^{(n)}\big(i_{P^{(n)}}(M,J,\mathbb{P}^{\text{emp}}_{X^n,\mathbf{v}^{(n)}(M,J)})\geq n(R+R_c+\varepsilon_1)\big)
}\nonumber\\*
= \ &P^{(n)}\big(P^{(n)}(M,J,\mathbb{P}^{\text{emp}}_{X^n,\mathbf{v}^{(n)}(M,J)})\leq 2^{-n(R+R_c+\varepsilon_1)}\big)&\nonumber\\
\leq \ &\lfloor 2^{nR} \rfloor \cdot \lfloor 2^{nR_c} \rfloor \cdot n^{|\mathcal{X}||\mathcal{V}|} \cdot 2^{-n(R+R_c+\varepsilon_1)} \underset{n\to\infty}{\longrightarrow} 0.\qquad \quad&\nonumber
\end{IEEEeqnarray}Since $J$ is uniformly distributed and independent of $X^n,$ and $P^{(n)}_{X^n} \equiv p_X^{\otimes n},$ then for any $(x^n,j)$ we have $i_{P^{(n)}}(x^n,j)=i_{p_X^{\otimes n}}(x^n)+\log(\lfloor 2^{nR_c} \rfloor).$ From the law of large numbers:
\begin{IEEEeqnarray}{c}
P^{(n)}(i_{P^{(n)}}(X^n,J)\leq n(H(X)+R_c-\varepsilon_2)) \underset{n\to\infty}{\longrightarrow} 0.\nonumber
\end{IEEEeqnarray}
When both the above events do not hold, we have
\begin{IEEEeqnarray}{c}
i_P(M,J)-i_P(X^n,J) \leq -n(H(X)-R-\varepsilon_1-\varepsilon_2).\nonumber
\end{IEEEeqnarray}
Hence, \eqref{eq:assumption_in_soft_covering_general_source_and_channel} holds.\\

\noindent \underline{Conclusion}
\hfill\\
For each $n{\geq}1,$ choose a realization of $\mathcal{B}^{(n)}$ giving a total variation distance below average, which defines a deterministic
\begin{IEEEeqnarray}{c}
f^{(n)}: (x^n,j) \mapsto \psi^{(n)}(x^n,j,u^{(n)}(x^n,j)) \in [2^{nR}].
\end{IEEEeqnarray}Moreover, for every $n{\geq}1,$ $\Tilde{\Phi}^{(n)}$ defines the same distribution of inputs $(X^n,J)$ as do the distributions $\Tilde{P}^{(n)}$ of Proposition \ref{prop:removing_encoder_private_randomness}.
We conclude using Lemma \ref{lemma:TV_joint_to_TV_marginal}, and the fact that the empirical distribution
is bounded and the average empirical distribution is its expectation, because $\mathcal{X}\times\mathcal{V}$ is finite.

\section{Converse of Theorem \ref{thm:region_discrete_alphabets}
}\label{app:converse_discrete_alphabets}

By definition, for each of the three statements of Theorem \ref{thm:region_discrete_alphabets}, we know that \scalebox{0.9}{$\mathcal{A}^{(f)}_D,\mathcal{A}^{(f,*)}_D,\mathcal{A}^{(u,*)}_D$} are included in \scalebox{0.9}{$\mathcal{A}^{(u)}_D.$} Therefore,
we
only
prove that
\scalebox{0.9}{$\mathcal{A}^{(u)}_D {\subseteq}
\mathcal{S}_D,$} in the setting of the first statement,
and
this also implies the same relation in the respective settings of the two other statements.
This proof closely tracks that of  \cite[Section~VI]{2013PaulCuffDistributedChannelSynthesis} and the end of \cite[Appendix~2]{2013PaulCuffDistributedChannelSynthesis}.
Let \scalebox{0.9}{$(R,R_c,R_d,\Delta) \in [0,H(X)) \times \overline{\mathbb{R}}_{\geq 0}^3$} be achievable with near-perfect realism with variable-length codes.
Fix \scalebox{0.9}{$\varepsilon$\hspace{2pt}$\in$\hspace{2pt}$(0,1/4).$}
Then,
for $n$ large enough,
there exists a \scalebox{0.9}{$(n,R,R_c,R_d)$} code inducing a joint distribution \scalebox{0.9}{$P_{X^n,J,M,Y^n}$} such that
\begin{IEEEeqnarray}{c}
\mathbb{E}_P[d(X^n, Y^n)] \, {\leq} \, \Delta \, {+} \, \varepsilon, \ 
\| P_{Y^n} {-} p_X^{\otimes n}\|_{TV} \leq \varepsilon \nonumber\\
\scalebox{0.92}{$H(M) \leq n
R
, \ 
H(J) \leq n
R_c
, \
H(L_d) \leq n
R_d
.$}\nonumber
\end{IEEEeqnarray}
Following the proof of \cite[Theorem~2]{2022AaronWagnerRDPTradeoffTheRoleOfCommonRandomness} and \cite[Section~VI~\&~Appendix~2]{2013PaulCuffDistributedChannelSynthesis}, we have the following claim.
\vspace{5pt}
\begin{claim}\label{claim:rate_lower_bounds_in_converse}
By introducing a random index uniformly distributed on $[n],$ one can construct a distribution \scalebox{0.9}{$P^{(\varepsilon)}_{X,Y,V}$} on \scalebox{0.9}{$\mathcal{X}^2\times\mathcal{V}$} satisfying \scalebox{0.9}{$X{-}V{-}Y, \ \| P_Y {-} p_X\|_{TV} {\leq} \varepsilon,$} and rate bounds
\begin{IEEEeqnarray}{c}
\scalebox{0.9}{$ \
R
\geq I(X;V), \
R{+}R_c
\geq I(Y;V){-}g(\varepsilon), \
R_d
\geq H(Y|V),$}\nonumber
\end{IEEEeqnarray}for some deterministic function $g$ such that $g(\varepsilon) {\to} 0$ as $\varepsilon {\to} 0.$
\end{claim}
\indent See Appendix \ref{app:rate_lower_bounds} for a proof.
We can change $V$ so that $|\mathcal{V}| {\leq} |\mathcal{X}|^2{+}1$ while preserving $P^{(\varepsilon)}_{X,Y},$ the Markov chain $X{-}V{-}Y,$ and quantities $I(X;V),I(Y;V)$ (and thus $H(Y|V)
$).

This follows from the proof of \cite[Lemma~VI.1]{2013PaulCuffDistributedChannelSynthesis}.
We then conclude with the same argument as in \cite[Section~VI]{2013PaulCuffDistributedChannelSynthesis}, as follows.
Consider a vanishing sequence $\{\varepsilon_n\}_{n \geq 1}$ in $(0,1/4).$ Owing to the cardinality bound, all probabilities $P^{(\varepsilon_n)}$ can be considered as points in the compact standard simplex of $\mathbb{R}^{2|\mathcal{X}|{+}|\mathcal{X}|^2{+}1}.$ Hence, a sub-sequence converges towards some probability $P^*_{X,V,Y}$ in the latter.
This distribution satisfies the Markov chain constraint and $P^*_X {\equiv} p_X {\equiv} P^*_Y.$ Hence, $P^* {\in} \mathcal{D}_D.$ Moreover,
since $R{<}H(X)$ and $g(\varepsilon_n){\to} 0$ as $n {\to} \infty,$ and
from
the rate lower bounds in Claim \ref{claim:rate_lower_bounds_in_converse},
we have
$(R,R_c,R_d,\Delta) {\in} \mathcal{S}_D,$ which concludes the converse proof.

\section{Rate lower bounds in the converse proof}\label{app:rate_lower_bounds}
Here, we provide a proof of Claim \ref{claim:rate_lower_bounds_in_converse} in Appendix \ref{app:converse_discrete_alphabets}.
Let $T$ be a uniform random variable on $[n].$ 
Define $V\text{=}(M, J, T, Y^{T-1}).$ 
Since $X^n \sim p_X^{\otimes n},$ the distribution of $X_T$ is $p_X.$ 
From Lemma \ref{lemma:TV_same_channel},
we have $\| P_{Y_T} {-} p_X\|_{TV} \leq \varepsilon.$
From the Markov chain $X^n-(M,J)-Y^n,$ the distribution $P_{X_T,V,Y_T}$ satisfies the
Markov chain
$X_T-V-Y_T.$

\noindent Since $P_{X_T,Y_T} \equiv \hat{P}_{\mathcal{X}^2}[X^n,Y^n],$ we have $\mathbb{E}[\scalebox{1.0}{$d(X_T, Y_T)$}] = \mathbb{E}[\scalebox{1.0}{$d(X^n, Y^n)$}] \leq \Delta + \varepsilon.$
We derive rate lower bounds using the following lemma.
\vspace{5pt}
\begin{lemma}\label{lemma:nearly_iid}\cite[Lemma~VI.3]{2013PaulCuffDistributedChannelSynthesis}
For any finite alphabet $\mathcal{W}$ and any random sequence $\Pi_{W^n}$ taking values in $\mathcal{W}^n,$ if there exists a distribution $\Gamma_W$ on $\mathcal{W}$ such that
\begin{IEEEeqnarray}{c}
\|\Pi_{W^n} - \Gamma_W^{\otimes n}\|_{TV} \leq \varepsilon< \dfrac{1}{4}, \quad \text{then,}\nonumber
\\
\dfrac{1}{n} \sum_{t=1}^n I_{\Pi}(W_t;W^{t-1}) \leq 4\varepsilon \Big( \log(|\mathcal{W}|)+ \log\Big(\dfrac{1}{\varepsilon}\Big) \Big),\nonumber
\end{IEEEeqnarray}and for any random variable $T$ uniformly distributed on $[n]$ and independent of $W^n,$ we have
\begin{IEEEeqnarray}{c}
I_{\Pi}(W_T;T) \leq 4\varepsilon \Big( \log(|\mathcal{W}|)+ \log\Big(\dfrac{1}{\varepsilon}\Big) \Big).\nonumber
\end{IEEEeqnarray}
\end{lemma}
Define
\begin{IEEEeqnarray}{c}
g:(0,1/4) \to (0,\infty), \ \varepsilon \mapsto 4\varepsilon \Big( \log(|\mathcal{X}|)+ \log\Big(\dfrac{1}{\varepsilon}\Big) \Big).\nonumber
\end{IEEEeqnarray}
We have
\begin{IEEEeqnarray}{rCl}
    nR
    \geq H(M)
    &\geq& \scalebox{1.0}{$I(M;X^n|J)$} \nonumber \\
    &=& \scalebox{1.0}{$I(M, J ;X^n)$} \label{eq:converse_E_D_marginal_R_using_J_indep} \\
    &=& \scalebox{1.0}{$\sum_{t=1}^n I(M, J ; X_t | X_{t+1:n})$} \nonumber \\ 
    &=& \scalebox{1.0}{$\sum_{t=1}^n I(M, J, X_{t+1:n} ; X_t)$} \nonumber \\
    &\geq& \scalebox{1.0}{$\sum_{t=1}^n I(M, J ; X_t)$} \nonumber \\
    &=& \scalebox{1.0}{$\sum_{t=1}^n I(M, J, Y^{t-1} ; X_t)$} \label{eq:introducing_Y_in_V} \\
    &=& \scalebox{1.0}{$n I(M, J, Y^{T-1} ; X_T | T)$}\IEEEeqnarraynumspace\label{eq:converse_E_D_marginal_lower_bound_R_t_to_T} \\  
    &=& \scalebox{1.0}{$n I(V ; X_T),$}\label{eq:converse_E_D_marginal_lower_bound_R_T_indep}
\end{IEEEeqnarray}
where \eqref{eq:converse_E_D_marginal_R_using_J_indep} follows from the independence between the common randomness and the sources; and \eqref{eq:introducing_Y_in_V} follows from Markov chain $X^n-(M,J)-Y^n;$ and \eqref{eq:converse_E_D_marginal_lower_bound_R_t_to_T} and \eqref{eq:converse_E_D_marginal_lower_bound_R_T_indep} follow from the independence of $T$ and all other variables and from the fact that variables in $\{X_t\}_{t \in [n]}$ are i.i.d.. We also have
\begin{IEEEeqnarray}{rCl}
    n(R{+}R_c
    ) &\geq& I(M,J;Y^n) \nonumber \\*
    &=& \sum_{t=1}^n I(M,J;Y_t|Y^{t-1}) \nonumber \\*
    &=& \sum_{t=1}^n \big[ I(M,J,Y^{t-1} ;Y_t) - I(Y^{t-1} ;Y_t) \big] 
    \nonumber
    \\*
    &\geq& \sum_{t=1}^n I(M,J,Y^{t{-}1} ;Y_t) 
    -ng(\varepsilon)
    \IEEEeqnarraynumspace \label{eq:converse_R_R_0_using_lemma_3_of_Yassaee_et_al} 
    \\*
    &=& nI(M,J,Y^{T{-}1} ;Y_T|T)
    - ng(\varepsilon)
    \nonumber \\*
    &=& nI(T,M,J,Y^{T{-}1} ;Y_T)
    {-} nI(T;Y_T) {-} ng(\varepsilon)
    \nonumber
    \\*
    &\geq& nI(V;Y_T)
    -2ng(\varepsilon)
    ,\label{eq:converse_R_R_0_using_lemma_4_of_Yassaee_et_al}\IEEEeqnarraynumspace
\end{IEEEeqnarray}where \eqref{eq:converse_R_R_0_using_lemma_3_of_Yassaee_et_al} and \eqref{eq:converse_R_R_0_using_lemma_4_of_Yassaee_et_al} follow from
Lemma \ref{lemma:nearly_iid}.
Moreover,
\begin{IEEEeqnarray}{rCl}
    nR_d
    \geq H(L_d)
    \geq I(L_d;Y^n|M,J) 
    &=& H(Y^n|M,J) \nonumber \\*
    &=& \sum_{t=1}^n H(Y_t|Y^{t-1},M,J) \nonumber \\*
    &=& nH(Y_T|V).
   \nonumber
\end{IEEEeqnarray}

\section{Proof of Theorem \ref{thm:region_general_alphabets}}\label{app:proof_coding_thm_general}
\subsection{Converse}
We know that $\mathcal{A}^{(f,*)}_G {\subseteq} \mathcal{A}^{(f)}_G.$ Moreover, the inclusion $\mathcal{A}^{(f)}_G \subseteq \overline{\mathcal{S}_G}$ is
the converse direction of \cite[Theorem~2]{2022AaronWagnerRDPTradeoffTheRoleOfCommonRandomness}, which is indeed stated for fixed-length codes \cite[Definition~1]{2022AaronWagnerRDPTradeoffTheRoleOfCommonRandomness}.


\subsection{Quantization argument}

Let $(R,R_c,\Delta)$ be a triplet in $\mathcal{S}_{G}.$ Let $p_{X,Y,V}$ be a corresponding distribution from the definition of $\mathcal{S}_{G}.$ Then
\begin{equation}\label{eq:introducing_R_R_c_Delta_general}
    \scalebox{0.9}{$R \geq I_p(X;V), \ R+R_c \geq I_p(Y;V), \ \Delta \geq \mathbb{E}_p[d(X,Y)].$}
\end{equation}

\noindent Fix some $\varepsilon>0.$
By assumption, $(d,p_X)$ is quantizable and uniformly integrable. Let $\tau'$ be a threshold corresponding to $\varepsilon$ as in Definition \ref{def:uniform_integrability}.
Set $\tau = \tau'/2.$ Let $\{\kappa^{(\ell)}\}_{\ell \in \mathbb{N}},$ $B_\tau$ and $L_{\varepsilon,\tau}$ be as in Definition \ref{def:continuious_distortion_function}. 
The former is a sequence of measurable quantizers of $\mathcal{X}$ such that the corresponding partitions asymptotically generate its Borel $\sigma$-algebra -i.e. quantization becomes arbitrarily fine as $\ell$ grows.
Therefore,
ince the source has infinite entropy, then from \cite[Lemma~7.18]{2011GrayGeneralAlphabetsRelativeEntropy}, there exists $L'_\varepsilon$
such that
for any $\ell \geq L'_{\varepsilon},$
\begin{IEEEeqnarray}{c}
\scalebox{0.9}{$H_p(\kappa^{(\ell)}(X))>R.$}\label{eq:R_less_than_entropy_quantized_X}
\end{IEEEeqnarray}
Fix $\ell \geq \max(L'_\varepsilon, L_{\varepsilon,\tau}).$
We denote $\kappa^{(\ell)}(X)$ by $[X]$ and $\kappa^{(\ell)}(Y)$ by $[Y].$
Since $p$ satisfies $X-V-Y$ and $p_Y \equiv p_X,$ then we have $[X]-V-[Y]$ and $p_{[Y] \equiv p_{[X]}}.$ Thus,
\begin{IEEEeqnarray}{c}
p_{[X],V,[Y]} \in \mathcal{D}_D,\label{eq:p_in_D_D_quantized}
\end{IEEEeqnarray}where $\mathcal{D}_D$ is the set defined in 
\eqref{eq:def_D_Discrete}, corresponding to source distribution $p_{[X]}$ (instead of $p_X$).
\noindent From \eqref{eq:in_def_quantizability_proba} and a union bound, we have
\begin{IEEEeqnarray}{c}
\scalebox{0.95}{$p\big((X,Y) \notin B_\tau^2\big) \leq p(X\notin B_\tau) + p(Y\notin B_\tau) \leq 2\tau=\tau',$}\IEEEeqnarraynumspace\label{eq:proba_of_being_outside}
\end{IEEEeqnarray}then from the uniform integrability we have
\begin{equation}
        0 \leq \mathbb{E}_p[d(X,Y) \mathbf{1}_{(X,Y) \notin B_\tau^2}] \leq \varepsilon.\label{eq:last_quantization_inequality_two_sided}
\end{equation}
\noindent From \eqref{eq:in_def_quantizability_distortion},
for every
$\ell \geq L_{\varepsilon,\tau}$ and every 
$(x,y)\in B_\tau^2,$ we have
\begin{IEEEeqnarray}{c}
\big| d(x,y) - d\big([x]), [y]\big)\big| \leq \varepsilon,
\label{eq:distortion_effect_of_quantization_two_sided}
\end{IEEEeqnarray}
and for every 
$(x,y)\in \mathcal{X}^2 \setminus B_\tau^2,$ we have
\begin{IEEEeqnarray}{c}
d\big([x]), [y]\big) \leq d(x,y) +\varepsilon.
\label{eq:dealing_with_the_outside_two_sided}
\end{IEEEeqnarray}
From \eqref{eq:distortion_effect_of_quantization_two_sided} and \eqref{eq:dealing_with_the_outside_two_sided},
we get
\begin{equation}\label{eq:initial_ineq_distortion_with_quantizations_two_sided}
    \mathbb{E}_p[d(X,Y)] + \varepsilon
    \geq \mathbb{E}_p[d([X],[Y])].
\end{equation} 
\noindent Since $\kappa^{(\ell)}$ is deterministic,
then
from \eqref{eq:introducing_R_R_c_Delta_general} and \eqref{eq:initial_ineq_distortion_with_quantizations_two_sided},
we have
\begin{IEEEeqnarray}{c}
\scalebox{0.95}{$R {\geq} I_p([X];V), \ R {+} R_c {\geq} I_p([Y];V), \ \Delta {+} \varepsilon
{\geq} \mathbb{E}[d([X],[Y])].$}\IEEEeqnarraynumspace\label{eq:initial_ineq_R_R_c_Delta_with_quantizations}
\end{IEEEeqnarray}
\noindent Let $\varepsilon' \in (0,\varepsilon]$ such that
\begin{IEEEeqnarray}{c}
    R+\varepsilon' < H_p([X]).\label{eq:rate_less_than_conditional_entropy}
\end{IEEEeqnarray}
\noindent Hence, from \eqref{eq:p_in_D_D_quantized} and since the auxiliary variable $V$ is assumed to be finite-valued, we can apply Proposition \ref{prop:can_construct_non_privately_randomized_discrete_code} with distribution $p_{[X],V,[Y]}$ and triplet $(R+\varepsilon',R_c,\Delta+\varepsilon
).$ Hereafter, we use the notation of Proposition \ref{prop:can_construct_non_privately_randomized_discrete_code}.


\subsection{Transition from near-perfect to perfect realism}\label{app:subsec:equivalence_perfect_realism}

\begin{proposition}\label{prop:to_perfect_realism_plain}
Let $n$ be a positive integer and $\delta$ be 
 a positive real.
Let $\mathcal{X}$
and $\mathcal{U}$ be two 
Polish alphabets and $p_{X}$ be a distribution on $\mathcal{X}.$ 
Let $d$ be a distortion measure such that $(d,p_X)$ is uniformly integrable.
Let $P_{X^n,U,Y^n}$ be a distribution on $\mathcal{X}^n\times\mathcal{U}\times\mathcal{X}^n$
and Markov chain property $X^n-U-Y^n.$
Moreover,
assume that
\begin{equation}
    \| P_{Y^n} - p_X^{\otimes n} \|_{TV} \leq \delta.\label{eq:total_variation_Q_P_on_Y_assumption_theorem_to_perfect_marginal_realism}
\end{equation}
Then, there
exists a conditional distribution $P'_{Y^n|U}$
such that the distribution $P'$ defined by 
\begin{IEEEeqnarray}{c}
P'_{X^n,U,Y^n} := P_{X^n,U} \cdot P'_{Y^n|U}
\end{IEEEeqnarray}
satisfies
\begin{IEEEeqnarray}{c}
\|P'_{X^n,U^n,Y^n}-P_{X^n,U^n,Y^n}\|_{TV} \leq \delta
\label{eq:TV_P_tweaked_P_original}\\*
\text{ and } 
P'_{Y^n} \equiv p_X^{\otimes n}.\label{eq:perception_result_theorem_to_perfect_realism}
\end{IEEEeqnarray}
\end{proposition}
\vspace{5pt}
\begin{IEEEproof}
This is a simple reformulation of the construction laid out in the proof of \cite[Theorem~1]{2022AaronWagnerRDPTradeoffTheRoleOfCommonRandomness}. Our variable $U$ is the tuple $(J,I=F_n(X^n,J))$ therein, which satisfies Markov chain $X^n-(J,F_n(X^n,J))-Y^n$ by \cite[Definition~2]{2022AaronWagnerRDPTradeoffTheRoleOfCommonRandomness}. Nothing in the proof in \cite{2022AaronWagnerRDPTradeoffTheRoleOfCommonRandomness} truly relies on any property of $(J,I).$ In particular, then uniform distribution of $J$ therein can be replaced by any distribution. Our distributions $p,P,P'$ are distributions $P,$ $PW,$ and $P\Tilde{W}$ in \cite{2022AaronWagnerRDPTradeoffTheRoleOfCommonRandomness}, respectively. Moreover, \eqref{eq:perception_result_theorem_to_perfect_realism} is \cite[(26)]{2022AaronWagnerRDPTradeoffTheRoleOfCommonRandomness} and \eqref{eq:TV_P_tweaked_P_original} follows from \cite[(27),(30),(39)]{2022AaronWagnerRDPTradeoffTheRoleOfCommonRandomness}.
\end{IEEEproof}


\noindent For each $n \in \mathbb{N},$ we define
\begin{IEEEeqnarray}{c}
\delta_1^{(n)}{:=}\|P^{(n)}_{[Y]^n}{-}p_X^{\otimes n}\|_{TV},
\nonumber
\end{IEEEeqnarray}
and apply Proposition \ref{prop:to_perfect_realism_plain} to $P^{(n)}$ with single-letter distribution $p_{[X]}$ and with $U{=}(M,J),\delta
{=}
\delta_1^{(n)}
.$ We denote the resulting distribution by ${P'}^{(n)}.$
From the triangle inequality for the total variation distance, Lemma \ref{lemma:TV_joint_to_TV_marginal}, and \eqref{eq:TV_P_tweaked_P_original}, we get
\begin{IEEEeqnarray}{c}
\Big\|\hat{P'}^{(n)}_{[\mathcal{X}]^2}[[X]^n,[Y]^n] - \hat{P}^{(n)}_{[\mathcal{X}]^2}[[X]^n,[Y]^n]\Big\|_{TV} \leq \delta_1^{(n)}.\label{eq:TV_empirical_P_tweaked_P_original}
\end{IEEEeqnarray}
From the additivity of $d,$ we have
\begin{IEEEeqnarray}{c}
\mathbb{E}_{P^{(n)}}[d([X]^n,[Y]^n)] = \mathbb{E}_{\hat{P}^{(n)}_{[\mathcal{X}]^2}[[X]^n,[Y]^n]}[d([X],[Y])].
\nonumber
\end{IEEEeqnarray}
Since $d$
does not take infinite values (Definition \ref{def:distortion})
and
$[\mathcal{X}]$ is finite, then $d$ is bounded on $[\mathcal{X}]^2.$
Hence, from Lemma 
\ref{lemma:continuity_TV}, \eqref{eq:distortion_on_P_we_get_from_lemma_discrete_code}, and \eqref{eq:TV_empirical_P_tweaked_P_original}, we get
\begin{IEEEeqnarray}{c}
\mathbb{E}_{\hat{P'}^{(n)}_{\mathcal{X}^2}[X^n,Y^n]}[d(X,Y)] \leq \mathbb{E}_{P^{(n)}}[d([X]^n,[Y]^n)] + \delta^{(n)}_2,
\label{eq:distortion_error_from_tweaking}
\IEEEeqnarraynumspace
\end{IEEEeqnarray}where $\delta^{(n)}_2 =2\max(d)\delta_1^{(n)},$
so that
$
\delta^{(n)}_2
\to
0.
$
For any $n {\in} \mathbb{N},$
the following distribution
\begin{IEEEeqnarray}{rCl}
\scalebox{0.9}{$\Tilde{P}^{(n)}_{X^n,[X]^n,J,M,[Y]^n,Y^n}$}&:=& p_X^{\otimes n} \cdot \prod_{t{=}1}^n \mathbf{1}_{\scalebox{0.65}{$[X]_t{=}[X_t]$}} \scalebox{0.9}{$\cdot p^{\mathcal{U}}_{[2^{nR_c}]} \cdot {P'}^{(n)}_{M|[X]^n,J}$}
\nonumber\\*
&\cdot& {P'}^{(n)}_{[Y]^n|M,J}
\cdot \prod_{t{=}1}^n p_{\scalebox{0.7}{$X|[X]{=}[Y]_t$}},\nonumber
\end{IEEEeqnarray}defines a \scalebox{0.9}{$(n,R+\varepsilon'+\varepsilon_n,R_c,\infty)$} fixed-length code with non-privately randomized encoding satisfying perfect realism, where \scalebox{0.9}{$[X]^n,[Y]^n$} denote discrete variables, and \scalebox{0.9}{$[X_t]=\kappa^{(\ell)}(X_t).$}
Denote
\begin{IEEEeqnarray}{c}
\pi_{X,Y,[X],[Y]} := \hat{\Tilde{P}}^{(n)}_{\mathcal{X}^2\times[\mathcal{X}]^2}[X^n,Y^n,[X]^n,[Y]^n].
\end{IEEEeqnarray}
Then, we
have 
\begin{IEEEeqnarray}{c}
\pi_{[X],[Y]} \equiv \hat{P'}^{(n)}_{[\mathcal{X}]^2}[[X]^n,[Y]^n]
\quad \text{and} \quad
\pi_Y \equiv \pi_X \equiv p_X.
\label{eq:Tilde_P_same_quantized_as_P'}
\IEEEeqnarraynumspace
\end{IEEEeqnarray}


\subsection{Conclusion}

We have
\begin{IEEEeqnarray}{rCl}
\IEEEeqnarraymulticol{3}{l}{
\mathbb{E}_{\Tilde{P}^{(n)}}[d(X^n, Y^n)] 
}\nonumber\\*
&=& \mathbb{E}_{\hat{\Tilde{P}}^{(n)}_{\mathcal{X}^2}[X^n,Y^n]}[d(X, Y)]\label{eq:using_additivity}\\*
&=& \mathbb{E}_\pi[d(X, Y)]\nonumber\\*
&=& \mathbb{E}_\pi\Big[d(X,Y)\substack{\scalebox{1.0}{$\mathbf{1} \qquad \quad \ $}  \\ (X,Y) \in B_\tau^2}
\Big]\nonumber\\*
&+& \mathbb{E}_\pi\Big[d(X, Y)\substack{\scalebox{1.0}{$\mathbf{1} \qquad \qquad \ $}  \\ (X,Y) \notin B_\tau^2}\Big] \nonumber\\*
&\leq& \mathbb{E}_\pi\Big[d([X],[Y])\substack{\scalebox{1.0}{$\mathbf{1} \qquad \qquad \ $}  \\ (X,Y) \in B_\tau^2}
\Big]+\varepsilon
+\varepsilon
\label{eq:using_quantizability}
\\
&\leq&
\mathbb{E}_{\hat{P'}^{(n)}_{[\mathcal{X}]^2}[[X]^n,[Y]^n]}[d([X], [Y])]
2\varepsilon
\nonumber\\
&\leq& \Delta+3\varepsilon
,\label{eq:final_final_distortion_bound}
\end{IEEEeqnarray} 
where
\eqref{eq:using_additivity} follows from the additivity of $d;$
\eqref{eq:using_quantizability} follows from \eqref{eq:distortion_effect_of_quantization_two_sided} and \eqref{eq:last_quantization_inequality_two_sided}; and \eqref{eq:final_final_distortion_bound}
holds for large enough $n$ from \eqref{eq:distortion_on_P_we_get_from_lemma_discrete_code} and \eqref{eq:distortion_error_from_tweaking}.
For every $n \in \mathbb{N},$ distribution $\Tilde{P}^{(n)}$
defines a \scalebox{0.9}{$(n,R+\varepsilon'+\varepsilon_n,R_c,\infty)$} fixed-length code with non-privately randomized encoding satisfying perfect realism.
From
the formulation of Proposition \ref{prop:can_construct_a_discrete_code},
we have
$\delta^{(n)}_{1,2},\varepsilon_n{\to} 0$ as $n {\to} \infty.$ Then, from \eqref{eq:final_final_distortion_bound}, and since
$\varepsilon' {\leq} \varepsilon,$ tuple $(R{+}2\varepsilon,R_c, \Delta{+}
3
\varepsilon
)$ is achievable with perfect realism with fixed-length codes with non-privately randomized encoding. 
This being true for every $\varepsilon{>}0,$
we get $(R, R_c, \Delta) \in \overline{\mathcal{A}_{G}},$ which concludes the proof.

\section{Proof of Theorem \ref{thm:region_per_symbol_realism_all_alphabets}}\label{app:proof_per_symbol_realism}
\subsection{Converse - finite source alphabet}
It is sufficient to use the same proof as that of the converse of Theorem \ref{thm:region_discrete_alphabets}. Indeed, in the latter, it can be checked that we only ever need information about the joint distribution of the symbols in $Y^n$ to lower bound $R+R_c.$

\subsection{Converse - source with infinite entropy}
It is sufficient to use the same proof as that of the converse of \cite[Theorem~2]{2022AaronWagnerRDPTradeoffTheRoleOfCommonRandomness}, except that
Proposition \ref{prop:to_perfect_realism_plain}
should be applied
to $(X_T,Y_T)$ rather than $(X^n,Y^n),$ where $T$ denotes a random index uniformly distributed on $[n].$

\subsection{Achievability - finite source alphabet}
Let $p_{X,V,Y} \in \mathcal{D}_D$ and $(R,R_d,\Delta)$ such that
\begin{IEEEeqnarray}{c}
\scalebox{0.9}{$R \geq I_p(X;V), \
R_d \geq H_p(Y|V), \
\Delta \geq \mathbb{E}_p[d(X,Y)].$}\nonumber
\end{IEEEeqnarray}
\begin{lemma}\label{lemma:Jun_Chen_et_al}\cite[Lemma~2,~Appendix~D]{Dec2022WeakAndStrongPerceptionConstraintsAndRandomness}
Let $\mathcal{X}$ and $\mathcal{V}$ be two Polish alphabets, $p_{X,V}$ be a distribution on $\mathcal{X}\times\mathcal{V},$ and $R>I_p(X;V).$ Then, there exists $K \in [2^{nR}]$ and a codebook
\begin{IEEEeqnarray}{c}
\scalebox{0.93}{$(v^n(m))_{m{\in}[K]} \subseteq \{v^n{\in}\mathcal{V}^n|\forall v'{\in}\mathcal{V}, |\mathbb{P}^{\text{emp}}_{v^n}(v') {-} p_V(v')| \leq \delta p_V(v')\},$}\nonumber
\end{IEEEeqnarray}
such that
\scalebox{0.9}{$\|Q_{X^n} - p_X^{\otimes n}\|_{TV} \underset{n \to \infty}{\longrightarrow}0,$}
where
\begin{IEEEeqnarray}{c}
Q_{M,V^n,X^n}:= p^{\mathcal{U}}_{[K]} \cdot \mathbf{1}_{V^n{=}v^n(m)} \cdot \prod_{t=1}^n p_{X|V{=}v_t(m)}.
\end{IEEEeqnarray}
\end{lemma}
Fix \scalebox{0.92}{$\delta>0.$} As shown in \cite[Appendix~D]{Dec2022WeakAndStrongPerceptionConstraintsAndRandomness}, from Lemma \ref{lemma:Jun_Chen_et_al}, there exists a sequence of conditional distributions \scalebox{0.9}{$(P_{V^n|X^n})_n$} from \scalebox{0.9}{$\mathcal{X}^n$} to the set of circular shifts of the above codewords, denoted $p_{\check{X}^n|X^n}$ therein, such that
\begin{IEEEeqnarray}{c}
\scalebox{0.92}{$\max_{t\in[n]} \|P_{X_t,V_t}{-}p_{X,V}\|_{TV} {\leq} \varepsilon_n {+} \varepsilon_\delta, \text{ with } \varepsilon_n {\underset{n\to\infty}{\longrightarrow}} 0,  \varepsilon_\delta {\underset{\delta\to 0}{\longrightarrow}} 0,$}\IEEEeqnarraynumspace\\*
\scalebox{0.93}{$\text{where }\forall n \in \mathbb{N}, \quad P_{X^n,V^n} := p_{X}^{\otimes n} \cdot P_{V^n|X^n}.$}
\end{IEEEeqnarray}
We use Lemma \ref{lemma:Jun_Chen_et_al} with a rate of $R+\delta,$ hence the set of circular shifts of the corresponding codewords is of size less than $2^{n(R+\delta)},$ for $n$ large enough. We simply send to the decoder the index of the codeword outputted by $P_{V^n|X^n}.$ The decoder then applies memoryless channel $p_{Y|V}.$ Thus,
\begin{IEEEeqnarray}{c}
\scalebox{0.92}{$\max_{t\in[n]} \|P_{X_t,V_t,Y_t}{-}p_{X,V,Y}\|_{TV} {\leq} \varepsilon_n {+} \varepsilon_\delta.$}\IEEEeqnarraynumspace
\end{IEEEeqnarray}
We then apply Proposition \ref{prop:removing_encoder_private_randomness} and Remark \ref{rem:memoryless_generator}, which yield a sequence 
$({P'}^{(n)})_n$ of $(n,R+\delta,0,\infty)$ fixed-length codes with non-privately randomized encoding with decoder $\prod p_{Y|V},$ satisfying
\begin{IEEEeqnarray}{c}
\scalebox{0.92}{$\max_{t\in[n]} \|{P'}^{(n)}_{X_t,V_t,Y_t}{-}p_{X,V,Y}\|_{TV} {\leq} \varepsilon_n {+} \varepsilon_\delta {+}\varepsilon'_n,$}\IEEEeqnarraynumspace
\end{IEEEeqnarray} for some vanishing $(\varepsilon'_n)_n.$ We know that each codeword (including circular shifts) has an empirical distribution close to $p_V$ in TVD. Hence, using the same argument as in the proof of Theorem \ref{thm:region_discrete_alphabets} (Appendix \ref{app:subsec:local_channel_synthesis}), we can simulate memoryless channel $p_{Y|V}$ with fixed-length private randomness of rate \scalebox{0.9}{$R_d{+}\delta,$} with asymptotically vanishing error in TVD, yielding a sequence 
\scalebox{0.9}{$({P''}^{(n)})_n$} of \scalebox{0.9}{$(n,R{+}\delta,0,R_d{+}\delta)$} fixed-length codes with non-privately randomized encoding, satisfying
\begin{IEEEeqnarray}{c}
\scalebox{0.92}{$\max_{t\in[n]} \|{P''}^{(n)}_{X_t,V_t,Y_t}{-}p_{X,V,Y}\|_{TV} {\leq} \varepsilon_n {+} \varepsilon_\delta {+}\varepsilon'_n{+}\varepsilon''_n.$}\IEEEeqnarraynumspace
\end{IEEEeqnarray}Since the alphabets are finite and the distortion does not take infinite values, this shows that $(R,R_d,\Delta)$ is in the closure of the set of triplets achievable with near-perfect per-symbol realism with no common randomness and no encoder private randomness, as desired.

\subsection{Achievability - source with infinite entropy}
It is sufficient to use the same proof as for Theorem \ref{thm:region_general_alphabets} (Appendix \ref{app:proof_coding_thm_general}) to go from a code on a quantized alphabet to a proper code. Indeed, Proposition \ref{prop:to_perfect_realism_plain} can be used with $n{=}1$ on each $([X_t],[Y_t])$ to perform the transition from near-perfect per-symbol realism to perfect per-symbol realism. This also yields an average empirical distribution of $Y^n$ equal to $p_X.$ Then, we can use the same
distortion bounds, because all TVD bounds are uniform in index $t,$ and imply a bound on the average empirical distribution of $([X]^n,[Y]^n).$

\section{Further justifications}

\subsection{Proof of Proposition \ref{prop:can_construct_a_discrete_code}}\label{app:subsec:proof_can_construct_a_discrete_code}
We start by proving the following result.
\vspace{5pt}
\begin{proposition}\label{prop:can_construct_a_discrete_code_with_Q}
Consider finite alphabets \scalebox{0.9}{$\mathcal{X},\mathcal{V},$} a distortion measure \scalebox{0.9}{$d$} on \scalebox{0.9}{$\mathcal{X}^2,$} a triplet \scalebox{0.9}{$(R,R_c,\Delta)\in\mathbb{R}_{\geq 0}^3,$} and a distribution \scalebox{0.9}{$p_{X,Y,V}$} on \scalebox{0.9}{$\mathcal{X}^2\times\mathcal{V}.$} Assume that \scalebox{0.9}{$p \in \mathcal{D}_D$} and
\begin{IEEEeqnarray}{c}
\scalebox{0.9}{$R \geq I_p(X;V), \
R+R_c \geq I_p(Y;V), \
\Delta \geq \mathbb{E}_p[d(X,Y)].$}\nonumber
\end{IEEEeqnarray}Then, there exists a sequence \scalebox{0.9}{$\{\varepsilon_n\}_{n \geq 1},$}
a sequence
of distributions
\scalebox{0.9}{$\{Q^{(n)}_{X^n,J,M,Y^n}\}_{n \geq 1}$}
and a sequence of \scalebox{0.9}{$(n,R+\varepsilon_n,R_c,\infty)$} fixed-length codes inducing distributions \scalebox{0.9}{$\{P^{(n)}_{X^n,J,M,Y^n}\}_{n \geq 1}$} such that
\scalebox{0.9}{$\varepsilon_n \underset{n \to \infty}{\rightarrow} 0$} and
\begin{IEEEeqnarray}{c}
\scalebox{0.9}{$\limsup_{n \to \infty} \mathbb{E}_{Q^{(n)}}[d(X^n,Y^n)] \leq \Delta$} \label{eq:distortion_on_Q_we_get_from_lemma_discrete_code_with_Q}\\
\scalebox{0.9}{$\|Q^{(n)}_{Y^n}-p_X^{\otimes n}\|_{TV} \underset{n \to \infty}{\longrightarrow} 0$}\label{eq:percetion_we_get_on_Q_from_lemma_discrete_code_with_Q}\\
\scalebox{0.9}{$\|P^{(n)}_{X^,J,M,Y^n} - Q^{(n)}_{X^,J,M,Y^n}\|_{TV} \underset{n \to \infty}{\longrightarrow} 0$}\label{eq:TV_P_Q_X_J_M_Y_from_lemma_discrete_code_with_Q}\\
\scalebox{0.9}{$\forall (m,j), \ P^{(n)}_{Y^n|M{=}m,J{=}j} = \prod_{t=1}^n p_{Y|V{=}v^{(n)}_t(m,j)},$}\label{eq:decoder_is_a_memoryless_generator_from_lemma_discrete_code_with_Q}\\
\scalebox{0.9}{$Q^{(n)}\big(\big\|\mathbb{P}^{\text{emp}}_{\mathcal{V}}(v^{(n)}(M,J)) - p_V\big\|_{TV} \geq \varepsilon_n\big) \underset{n \to \infty}{\longrightarrow} 0,$}\label{eq:codeword_letter_frequency_converges_in_proba_from_lemma_discrete_code_with_Q}
\end{IEEEeqnarray}for some sequence of deterministic functions \scalebox{0.9}{$v^{(n)}: [2^{n(R+\varepsilon)}]{\times}[2^{nR_c}] \to \mathcal{V}^{n}.$}
\end{proposition}
\vspace{5pt}
\begin{IEEEproof}
Except for \eqref{eq:codeword_letter_frequency_converges_in_proba_from_lemma_discrete_code_with_Q}, the above result follows directly from the random coding proof in \cite{2022AaronWagnerRDPTradeoffTheRoleOfCommonRandomness}, which tracks the achievability proof in \cite{2013PaulCuffDistributedChannelSynthesis}: $Q^{(n)}$ is $Q$ \cite[Eq~76]{2022AaronWagnerRDPTradeoffTheRoleOfCommonRandomness}, $P^{(n)}$ is $\Tilde{Q}$ \cite[Eq~82]{2022AaronWagnerRDPTradeoffTheRoleOfCommonRandomness}, \eqref{eq:distortion_on_Q_we_get_from_lemma_discrete_code_with_Q} is \cite[Eq.~79]{2022AaronWagnerRDPTradeoffTheRoleOfCommonRandomness}, \eqref{eq:percetion_we_get_on_Q_from_lemma_discrete_code_with_Q} is \cite[Eq.~77]{2022AaronWagnerRDPTradeoffTheRoleOfCommonRandomness}, \eqref{eq:TV_P_Q_X_J_M_Y_from_lemma_discrete_code_with_Q} is \cite[Eq.~82]{2022AaronWagnerRDPTradeoffTheRoleOfCommonRandomness} and \eqref{eq:decoder_is_a_memoryless_generator_from_lemma_discrete_code_with_Q} is in \cite[Eq.~81]{2022AaronWagnerRDPTradeoffTheRoleOfCommonRandomness}.
Moreover,
one can readily impose \eqref{eq:codeword_letter_frequency_converges_in_proba_from_lemma_discrete_code_with_Q}
from the law of large numbers, because
$Q^{(n)}$
is constructed from a realization of a random codebook
having
i.i.d. codewords, each having i.i.d. symbols of distribution $p_V.$
\end{IEEEproof}
Proposition \ref{prop:can_construct_a_discrete_code} follows from Proposition \ref{prop:can_construct_a_discrete_code_with_Q}. Indeed,
from the triangle inequality for the total variation distance, we have
\begin{IEEEeqnarray}{rCl}
\IEEEeqnarraymulticol{3}{l}{
\|P^{(n)}_{Y^n}-p_X^{\otimes n}\|_{TV}
\leq \|P^{(n)}_{Y^n}-Q^{(n)}_{Y^n}\|_{TV} + \|Q^{(n)}_{Y^n}-p_X^{\otimes n}\|_{TV}
}
.\nonumber
\end{IEEEeqnarray}From \eqref{eq:TV_P_Q_X_J_M_Y_from_lemma_discrete_code_with_Q} and Lemma \ref{lemma:TV_joint_to_TV_marginal}, and from \eqref{eq:percetion_we_get_on_Q_from_lemma_discrete_code_with_Q}, the above right hand side goes to zero as $n$ goes to infinity, yielding \eqref{eq:percetion_we_get_on_P_from_lemma_discrete_code}.
From the additivity of $d,$ we have
\begin{IEEEeqnarray}{c}
\mathbb{E}_{Q^{(n)}}[d(X^n,Y^n)] = \mathbb{E}_{\hat{Q}^{(n)}_{\mathcal{X}^2}[X^n,Y^n]}[d(X,Y)],
\nonumber
\end{IEEEeqnarray}and the same for $P^{(n)}.$
Moreover, from the 
triangle inequality for the total variation distance, Lemma \ref{lemma:TV_joint_to_TV_marginal}, and \eqref{eq:TV_P_Q_X_J_M_Y_from_lemma_discrete_code_with_Q}, we have
\begin{IEEEeqnarray}{c}
\Big\|\hat{P}^{(n)}_{\mathcal{X}^2}[X^n,Y^n]-\hat{Q}^{(n)}_{\mathcal{X}^2}[X^n,Y^n]\Big\|_{TV} \underset{n \to \infty}{\longrightarrow} 0.\label{eq:TV_on_empirical_P_and_Q}
\end{IEEEeqnarray}
From Definition \ref{def:distortion}, $d$ does not take infinite values. Since $\mathcal{X}$ is assumed to be finite, then from Lemma \ref{lemma:continuity_TV}, \eqref{eq:distortion_on_Q_we_get_from_lemma_discrete_code_with_Q}, and \eqref{eq:TV_on_empirical_P_and_Q}, we get \eqref{eq:distortion_on_P_we_get_from_lemma_discrete_code}. Moreover, \eqref{eq:codeword_letter_frequency_converges_in_proba_from_lemma_discrete_code} follows from \eqref{eq:codeword_letter_frequency_converges_in_proba_from_lemma_discrete_code_with_Q} and \eqref{eq:TV_P_Q_X_J_M_Y_from_lemma_discrete_code_with_Q}.

\subsection{Local channel synthesis argument}\label{app:subsec:local_channel_synthesis}
We prove Claim \ref{claim:local_channel_synthesis}.
Consider a functional representation $Y=\psi(U,V)$ of $p_{Y|V}.$ Similarly to the end of \cite[Appendix~2]{2013PaulCuffDistributedChannelSynthesis}, we apply the local channel synthesis lemma \cite[Corollary~VII.6]{2013PaulCuffDistributedChannelSynthesis} with codebook distribution $P_U$ and channel $(v,u) \mapsto \psi(v,u).$ Then, for any $\gamma>0,$ there exists a vanishing sequence $(\delta_n)_n$ such that for any $r>\gamma$ and any $n \in \mathbb{N},$
there exists a mapping $\Tilde{P}_{Y^n|V^n}$ with a fixed-length private randomness of rate 
$r$
such that
\begin{IEEEeqnarray}{c}
\scalebox{0.9}{$\big\|\Tilde{P}_{Y^n|V^n{=}v^n} - \prod_{t=1}^n p_{Y|V=v_t}\big\|_{TV} \leq \delta_n,$}\label{eq:conlusion_of_local_channel_synthesis}
\end{IEEEeqnarray}
for any $v^n \in \mathcal{V}^n$ for which $p^{(v^n)}_{V,Y}:=P^{\text{emp}}_{v^n} \cdot p_{Y|V}$ satisfies $r-\gamma \geq H_{p^{(v^n)}}(Y|V).$
Fix some $\gamma>0.$
From \eqref{eq:codeword_letter_frequency_converges_in_proba_from_lemma_discrete_code} and the continuity of entropy, we have
\begin{IEEEeqnarray}{c}
\scalebox{0.9}{$P^{(n)}_{M,J}\big(H_p(Y|V)+\gamma \geq H_{p^{(v^n(M,J))}}(Y|V)\big) \underset{n\to\infty}{\longrightarrow} 1.$}\label{eq:convergence_of_entropies_in_proba}
\end{IEEEeqnarray}

\noindent Take $r=R_d+2\gamma.$
For \textit{good} sequences $v^n,$ i.e. those for which the above event holds true, we have upper bound \eqref{eq:conlusion_of_local_channel_synthesis} since $R_d \geq H_p(Y|V).$ For other sequences $v^n,$ the total variation distance is upper-bounded by $1.$ 
Then, from Lemma \ref{lemma:get_expectation_out_of_TV}, \eqref{eq:decoder_is_a_memoryless_generator_from_lemma_discrete_code} and \eqref{eq:convergence_of_entropies_in_proba},
we get
\begin{IEEEeqnarray}{c}
\scalebox{0.9}{$\big\|P^{(n)}_{v^n(M,J)} \cdot \Tilde{P}_{Y^n|V^n} - P^{(n)}_{v^n(M,J),Y^n}\big\|_{TV} 
\underset{n\to\infty}{\longrightarrow}0.$}\nonumber
\end{IEEEeqnarray}Since $P^{(n)}$ satisfies Markov chain $(X^n,M,J)-v^n(M,J)-Y^n,$ then from Lemma \ref{lemma:TV_same_channel} we get
\begin{IEEEeqnarray}{c}
\scalebox{0.9}{$\big\|P^{(n)}_{X^n,M,J,v^n(M,J)} \cdot \Tilde{P}_{Y^n|V^n} - P^{(n)}_{X^n,M,J,v^n(M,J),Y^n}\big\|_{TV} 
\underset{n\to\infty}{\longrightarrow}0.$}\nonumber
\end{IEEEeqnarray}

\subsection{Quantizability for Euclidean spaces}\label{app:subsec:quantizability_Euclidean_space}
We provide a proof of Claim \ref{claim:Euclidean_quantizable}.
Let $\mathcal{X}$
be a finite-dimensional real vector space,
with Euclidean distance
denoted by $d,$
let $p$ be a distribution on $\mathcal{X},$ and let $s,\varepsilon,\tau$ be positive reals. Since $\mathcal{X}$ is the union of all balls of integer radius centered at the origin, there exists one, denoted
$B_\tau,$
such that
\begin{IEEEeqnarray}{c}
p(\mathcal{X} \setminus B_\tau) \leq \tau. \label{eq:sigma_finite}
\end{IEEEeqnarray}
Fix an orthonormal basis of $\mathcal{X}.$
A \textit{rectangular prism} is a set of points of $\mathcal{X}$ whose coordinates lie in a product of bounded real intervals.
Fix $\ell \in \mathbb{N}.$ We define a quantizer $\kappa^{(\ell)}$ on $\mathcal{X}$ as follows.
Each coordinate axis can be partitioned into half-open intervals of length $\ell^{-s}.$
Products of such intervals are called \textit{basic rectangular prisms}. The latter form a partition of $\mathcal{X}.$
Let $B_\ell$ denote the closed ball of radius $\ell,$ centered at the origin. Then, the Euclidean projection onto $B_\ell$ of any point $x$ in $\mathcal{X}$ is uniquely defined as the element of $B_\ell$ having minimal Euclidean distance to $x.$
Map $\kappa^{(\ell)}$ is defined as:
\begin{itemize}
\item On each basic rectangular prism $\mathfrak{r}$ which is included in the interior of $B_\ell,$ define $\kappa^{(\ell)}$ as the constant mapping to the center of $\mathfrak{r}.$
\item
On
each
basic rectangular prism
having non-empty intersection with the border of $B_\ell,$
define $\kappa^{(\ell)}$ as the constant mapping to
an arbitrarily chosen
representative element.
Then, the image of each point on the border of $B_\ell$ is well-defined.
\item For any point $x$ in the remainder of $\mathcal{X},$ let $x'$ denote its Euclidean projection onto (the border of) $B_\ell.$ Define $\kappa^{(\ell)}(x)$ as
$\kappa^{(\ell)}(x').$
\end{itemize}
Then, $\kappa^{(\ell)}$ is finite-valued and measurable.
For any $x\in\mathcal{X},$ we denote $\kappa^{(\ell)}(x)$ by $[x].$
Let $\mathfrak{C}_\ell$ denote the set of basic rectangular prisms included in the interior of $B_\ell.$ Then, the set $\cup_{\ell \geq 1} \mathfrak{C}_\ell$ asymptotically generates the set of closed rectangular prisms (products of closed intervals).
Hence, it asymptotically generates the Borel $\sigma$-algebra of $\mathcal{X}.$
From the triangle inequality,
we have
\begin{IEEEeqnarray}{c}
\forall
(x,y,\Tilde{x},\Tilde{y}) \in \mathcal{X}^4,
d(x,y)-d(\Tilde{x},\Tilde{y}) \leq d(x,\Tilde{x}) + d(\Tilde{y},y).\label{eq:multiple_triangle_ineq}
\IEEEeqnarraynumspace
\end{IEEEeqnarray}
Denote by $\Tilde{B}_\ell$ the reunion of all basic rectangular prisms having non-empty intersection with $B_\ell.$ Its diameter is at most $2\ell+2\ell^{-s},$ hence at most $4\ell.$
Simple calculus implies that there exists a constant $\zeta_{s,\varepsilon}>0$ depending only on $s$ and $\varepsilon,$ such that for any
$(a,b)\in[0,4\ell]^2,$ we have
\begin{IEEEeqnarray}{c}
|a^s-b^s| \leq \varepsilon+ |a-b| (\zeta_{s,\varepsilon}+s(4\ell)^{s-1}).\label{eq:Lipschitz}
\end{IEEEeqnarray}
Hence, from \eqref{eq:multiple_triangle_ineq}, for any $(x,y) \in \Tilde{B}_\ell^2,$ we have
\begin{IEEEeqnarray}{rCl}
\IEEEeqnarraymulticol{3}{l}{
|d(x,y)^s {-} d([x],[y])^s|
}
\nonumber\\*
\leq \varepsilon + \big(d(x,[x])+d(y,[y])\big)\big(\zeta_{s,\varepsilon}{+}s(4\ell)^{s-1}\big).
\nonumber
\end{IEEEeqnarray}
By construction, $\kappa^{(\ell)}$ maps each element $x$ of $\Tilde{B}_\ell$ to an element $x''$
at Euclidean distance at most $\sqrt{\dim(\mathcal{X})}\ell^{-s}$ of $x.$
Hence,
\begin{IEEEeqnarray}{rCl}
\IEEEeqnarraymulticol{3}{l}{
\forall (x,y) \in \Tilde{B}_\ell^2,
\quad
|d(x,y)^s - d([x],[y])^s|
\leq 2\varepsilon,
}
\label{eq:final_quantizability_error_in_Tilde_B}
\end{IEEEeqnarray}
for large enough $\ell.$
Fix some $(x,y)\in \mathcal{X}^2 \setminus \Tilde{B}_\ell^2.$
Denote by $x'$ and $y'$ the Euclidean projections of $x$ and $y$ onto $B_\ell.$ Since the latter is convex and closed, we have $d(x',y') \leq d(x,y).$ We also have
$d([x],[x']),d([y],[y'])\leq 2\sqrt{\dim(\mathcal{X})}l^{-s}.$
Hence, from \eqref{eq:multiple_triangle_ineq} and \eqref{eq:Lipschitz}, for large enough $\ell,$ we have
\begin{IEEEeqnarray}{c}
d([x],[y])^s{-}d([x'],[y'])^s
\leq 2\varepsilon.
\nonumber
\end{IEEEeqnarray}
Combining this, $d(x',y') {\leq} d(x,y),$ and \eqref{eq:final_quantizability_error_in_Tilde_B} for $(x',y')$ gives:
\begin{IEEEeqnarray}{c}
\forall (x,y)\in \mathcal{X}^2 \setminus \Tilde{B}_\ell^2,
\quad
d([x],[y])^s \leq d(x,y)^s+ 4\varepsilon,\label{eq:distortion_is_reduced_by_quantization}
\end{IEEEeqnarray} for large enough $\ell.$
Moreover,
for large enough $\ell,$ we have $B_\tau \subset \Tilde{B}_\ell.$
Since \eqref{eq:sigma_finite}, \eqref{eq:final_quantizability_error_in_Tilde_B} and \eqref{eq:distortion_is_reduced_by_quantization} are true for any $\varepsilon,\tau>0$ for large enough $\ell,$ this concludes the proof.

\end{document}